\UseRawInputEncoding
\documentclass[12pt,technote, onecolumn, draft]{IEEEtran}

\usepackage{mathrsfs}
\usepackage{multirow}
\usepackage{amsfonts,amssymb,amsmath,amsthm,bm}
\usepackage{color}
\usepackage{cite}
\usepackage{blkarray}
\usepackage{changes}
\usepackage{tikz}
\usetikzlibrary{matrix}

\usepackage{framed}

\usepackage{amsmath}
\usepackage{mathtools}
\usepackage{array}
\usepackage{tikz}
\usetikzlibrary{matrix}

\newcolumntype{C}{>{$}c<{$}}

\newtheorem{theorem}{Theorem}
\newtheorem{example}{Example}
\newtheorem{definition}{Definition}
\newtheorem{remark}{Remark}
\newtheorem{corollary}{Corollary}
\newtheorem{lemma}{Lemma}
\newtheorem{proposition}{Proposition}

\newcommand\bovermat[2]{%
  \makebox[0pt][l]{$\smash{\overbrace{\phantom{%
    \begin{matrix}#2\end{matrix}}}^{\text{#1}}}$}#2}
\newcommand\partialphantom{\vphantom{\frac{\partial e_{P,M}}{\partial w_{1,1}}}}

\tikzset{arrowstyle/.style={draw=black, -Latex}}

\begin{document}

\title{New Lower Bounds for the Minimum Distance of Cyclic Codes and Applications to Locally Repairable Codes$^\dag$}
\author{Jing Qiu, Weijun Fang, Fang-Wei Fu
\IEEEcompsocitemizethanks{\IEEEcompsocthanksitem Jing Qiu and Fang-Wei Fu are with Chern Institute of Mathematics and LPMC, Nankai University, Tianjin 300071, China (Emails: jingqiu0202@163.com, fwfu@nankai.edu.cn).}
\IEEEcompsocitemizethanks{\IEEEcompsocthanksitem Weijun Fang is with Key Laboratory of Cryptologic Technology and Information Security, Ministry of Education, Shandong University, Qingdao, 266237, China, and School of Cyber Science and Technology, Shandong University, Qingdao, 266237, China, and Quancheng Laboratory, Jinan 250103, China  (Email: fwj@sdu.edu.cn).
}

\thanks{$^\dag$This research is supported in part by the National Key Research and Development Program of China under Grant Nos. 2022YFA1005000, 2022YFA1004900, 2021YFA1001000 and 2018YFA0704703, the National Natural Science Foundation of China under Grant Nos. 62201322, 12141108, 62371259, 61971243 and 12226336, the Natural Science Foundation of Shandong Province under Grant No. ZR2022QA031, the Fundamental Research Funds for the Central Universities, Nankai University, and the Nankai Zhide Foundation.} }

\maketitle

\begin{abstract}
Cyclic codes are an important class of linear codes.
Bounding the minimum distance of cyclic codes is a long-standing research topic in coding theory, and several well-known and basic results have been developed on this topic. Recently, locally repairable codes (LRCs) have attracted much attention due to their repair efficiency in large-scale distributed storage systems.
In this paper, by employing the singleton procedure technique, we first provide a sufficient condition for bounding the minimum distance of cyclic codes with typical defining sets. Secondly, by considering a specific case, we establish a connection between bounds for the minimum distance of cyclic codes and solutions to a system of inequalities. This connection leads to the derivation of new bounds, including some with general patterns. In particular, we provide three new bounds with general patterns, one of which serves as a generalization of the Betti-Sala bound.
Finally, we present a generalized lower bound for a special case and construct several families of $(2, \delta)$-LRCs with unbounded length and minimum distance $2\delta$. It turns out that these LRCs are distance-optimal, and their parameters are new. To the best of our knowledge, this work represents the first construction of distance-optimal $(r, \delta)$-LRCs with unbounded length and minimum distance exceeding $r+\delta-1$.
\end{abstract}


\section{Introduction}

\label{intro}

Due to the elegant algebraic structure and the efficiency both in encoding and decoding, cyclic codes play an important role both in theory and practice. However, it's hard to determine the exact value of the minimum distance of cyclic codes. Thus the estimation of the minimum distance of cyclic codes is a classical and important problem in coding theory.

To reduce the repair bandwidth in a massive reliable scale distributed storage system, the concept of locally repairable codes (LRCs) \cite{Gop12} emerged. The $i$-th code symbol of an $[n, k]$ linear code $\mathcal{C}$ is said to have locality $r$ if it can be recovered by accessing at most $r$ other symbols
in $\mathcal{C}$. If all the code symbols have $r$-locality, we call $\mathcal{C}$ an $r$-LRC. However, when multiple node failures occur, the original concept of locality may not work. Prakash et al.~\cite{Pra12} introduced the concept of $(r, \delta)$-locality of linear codes, where $\delta\geq 2$, which generalized the notion of $r$-locality. The $i$-th code symbol of $\mathcal{C}$ is said to have $(r, \delta)$-locality ($\delta \geq 2$), if there exists a subset $S_i \subset \{1, 2, \ldots, n\}$, which is termed the repair set of $i$-th code symbol, such that $i\in S_i$, $|S_i|\leq r+\delta-1$ and the punctured code $\mathcal{C}|_{S_i}$ has minimum distance $d(\mathcal{C}|_{S_i}) \geq \delta$. The code $\mathcal{C}$ is said to have $(r, \delta)$-locality or be an $(r,\delta)$-LRC if all the code symbols have $(r, \delta)$-locality. Note that $(r,\delta)$-locality reduces to $r$-locality when $\delta=2$, and we call a code LRC if it has $r$-locality or $(r,\delta)$-locality. A Singleton-like bound for the minimum distance of an $(r, \delta)$-LRC is given as follows~\cite{Pra12}:
\begin{equation}\label{eq:singleton}
d(\mathcal{C}) \leq  n-k- \left(\left\lceil \frac{k}{r}\right\rceil -1 \right)(\delta-1) +1.
\end{equation}
LRCs achieving bound~\eqref{eq:singleton} are called Singleton-optimal. As various works have pointed out, for example, see \cite{GXY19}, the bound \eqref{eq:singleton} is not achievable under certain parameter restrictions.
For an $[n, k, d]$ linear code $\mathcal{C}$, if there is no linear code with parameters $[n, k, d^{'} > d]$, we then call $\mathcal{C}$ a distance-optimal code. As we can see, an LRC which is Singleton-optimal is also distance-optimal.

\subsection{Known Results}

\textit{$1)$ Lower Bounds for The Minimum Distance of Cyclic Codes.}
There are several lower bounds for the minimum distance of cyclic codes, including, among others, the Bose-Chaudhuri-Hocquenghem (BCH), Hartmann-Tzeng (HT), Roos, and shift bounds (see \cite{Bose60},\cite{Hart72},\cite{Roo83}, \cite{Lin86}). They are usually based on special patterns in the defining set of the cyclic codes. In \cite{Betti06}, Betti and Sala presented a new lower bound based on a new pattern by utilizing the technique termed ``singleton procedure'' proposed in \cite{Pon03}.

\textit{$2)$ Constructions of Singleton-optimal LRCs via Cyclic and Contacyclic Codes.}
Over the past decade, cyclic codes and constacyclic codes have emerged as vital components in the constructions of Singleton-optimal locally repairable codes (LRCs).
Motivated by a breakthrough construction of Singleton-optimal LRCs via subcodes of Reed-Solomon codes in~\cite{Tamo14}, Tamo et al.~\cite{Tamo15, Tamo16} constructed a family of $q$-ary Singleton-optimal cyclic $r$-LRCs of length $q-1$ and its factors.
Then Singleton-optimal $(r, \delta)$-LRCs with length divide $q-1$ or $q+1$ were proposed by Chen et\ al.\cite{Chen18},\cite{Chen19} via cyclic and constacyclic codes.

After this, constructing optimal LRCs with long code length has attracted a lot of attention from researchers. Surprisingly, utilizing cyclic codes, Luo et\ al. in \cite{Luo19} found that the code length of Singleton-optimal LRCs can be unbounded, i.e., the code length $n$ is not constrained by the field size $q$ for the minimum distance $d=3$ or 4 and $r \geq d-1$.
Moreover, subsequent works in \cite{Fang20} and \cite{Sun19} have further expanded upon these findings. Singleton-optimal $q$-ary $(r,\delta)$-LRCs with unbounded length for $r\geq d-\delta+1$ and minimum distance not more than $2\delta$ were constructed by cyclic and constacyclic codes.

Subsequently, the next goal was to construct optimal $(r,\delta)$-LRCs with long code length and minimum distance larger than $2\delta$.
The first construction of $q$-ary Singleton-optimal cyclic $r$-LRC with length $n>q+1$ and $d\geq 5$ was proposed by Fang et al. \cite{Fang21}, where $r=2$.
The results were then generalized to $(2,\delta)$-LRCs in \cite{Qiu22}, while the code length $n=\frac{(\delta+1)(q + 1)}{2^t}$ for some integer $t\geq 0$, and the minimum distance is $2\delta+2$.

\subsection{Our Contributions}

The main contributions of this paper are summarized as follows.
\begin{itemize}
    \item[(1)] By utilizing the singleton procedure technique, we first present a criterion for determining the pseudo-rank of a square matrix whose elements are identical in the direction parallel to the main diagonal (see Lemma~\ref{lem:main}). We then propose our main theorem (see Theorem~\ref{thm:generalbound}), which provides a sufficient condition for bounding the minimum distance of a family of cyclic codes whose defining sets are of a typical type. Consequently, many new bounds with flexible forms can be derived.
  \item[(2)] We consider a specific case of the main theorem and establish a connection between bounds for the minimum distance of cyclic codes and solutions to a system of inequalities (see Corollary \ref{cor:bound0}). Specifically, any solution to the system of inequalities can yield a lower bound for the minimum distance of cyclic codes. Some new bounds with general forms can be derived from the system of inequalities (see Theorems~\ref{thm:explicitbound01}, \ref{thm:explicitbound02} and \ref{thm:explicitbound03}). In particular, Theorem~\ref{thm:explicitbound01} serves as a generalization of the Betti-Sala bound in \cite{Betti06}.
  \item[(3)] We provide an alternative bound, which is a generalization of Theorem~\ref{thm:generalbound} for the case of $s=1$. According to the new bound, we propose several families of distance-optimal $(2,\delta)$-LRC with unbounded length and minimum distance $2\delta$ from cyclic and constacyclic codes (see Theorems~\ref{thm:olrcq-1}, \ref{thm:olrcq+1deven} and \ref{thm:olrcq+1dodd}). To the best of our knowledge, this is the first work of constructing distance-optimal $(r,\delta)$-LRCs with unbounded length and minimum distance exceeding $r+\delta-1$. We compare the results of this paper with the prior literature in Table~\ref{tab1}.

\end{itemize}

\begin{table}[!htb]
\caption{Comparison of the constructions of optimal (consta)cyclic LRCs with $n\geq q+1$}\label{tab1}
\resizebox{\textwidth}{38mm}{
\begin{tabular}{|c|c|c|c|c|}
  \hline
  Distance & Locality & Length &Optimality & References \\ \hline
  $3$ & $r\geq 2, \delta=2$ & unbounded & Singleton-optimal  & \cite{Luo19}  \\ \hline
  $4$ & $r\geq 3, \delta=2$ & unbounded & Singleton-optimal& \cite{Luo19}  \\ \hline
  $\delta+1$ & $r\geq 2, \delta\geq 2$   & unbounded& Singleton-optimal  &  \cite{Fang20}  \\ \hline
  $\delta+2$ & $r\geq 3, \delta\geq 2$   & unbounded& Singleton-optimal  &  \cite{Fang20}  \\ \hline
  $2\delta$ & $r\geq \delta+1\geq 3$   & unbounded& Singleton-optimal  &  \cite{Fang20}  \\ \hline
  $\delta+2\epsilon$,$1\leq\epsilon \leq \frac{\delta}{2}$ & $r\geq 2\epsilon+1,\delta$ even   & unbounded& Singleton-optimal  &  \cite{Sun19}  \\ \hline
  $\delta+2\epsilon$,$1\leq\epsilon \leq \frac{\delta-1}{2}$ & $r\geq 2\epsilon+1,\delta$ odd   & unbounded& Singleton-optimal  &  \cite{Sun19}  \\ \hline
  $6$ & $r\geq 3,\delta=2$   & $3(q+1)$ or $\frac{3(q+1)}{2}$ & Singleton-optimal  &  \cite{Fang21}  \\ \hline
  $2\delta+2$ & $r\geq 3, \delta\geq 3$   & $n=\frac{(\delta+1)(q + 1)}{2^t}$  & Singleton-optimal  &  \cite{Qiu22}  \\ \hline

  $2\delta$ &  $r=2, \delta\geq 2$ & unbounded & distance-optimal & Theorems~\ref{thm:olrcq-1},\ref{thm:olrcq+1deven},\ref{thm:olrcq+1dodd} \\
  \hline
\end{tabular}}
\end{table}

\subsection{Organizations}
The rest of this paper is organized as follows. In Section II, we review some preliminaries on cyclic and constacyclic codes with their connections to LRCs and present an introduction to the technique of singleton procedure. In Section III, we present our main theorem for bounding the minimum distance of cyclic codes and provide some explicit bounds. In Section IV, we generalize our main theorem for a special case. In Section V, we present our constructions of distance-optimal LRCs by utilizing the results from Section IV. We use some conclusions to end this paper in Section VI.

\section{Preliminaries}
\label{sec:2}

Throughout this paper, we adopt the following notations unless otherwise stated:
\noindent\begin{itemize}
\item $q$ is a power of a prime, $\mathbb{F}_q$ is a finite field of size $q$, $\mathbb{F}_q^{*}=\mathbb{F}_q\setminus \{0\}$.
\item $n$ is a positive integer with $(n,q)=1$.
\item $\xi$ is a primitive $n$-th root of unity.
\item $\lambda$ is a nonzero element in $\mathbb{F}_{q}$, $\eta$ is in some extension field of $\mathbb{F}_q$ such that $\eta^{n}=\lambda$.
\item For any $n$-tuple $\mathbf{v}$ (including a vector in a linear space), let $M(\mathbf{v})$ denote the $n\times n$ circulant matrix obtained from $\mathbf{v}$, i.e., the matrix whose first row is $\mathbf{v}$ and whose other rows are obtained by cyclic shifting.
\item $[n]=\{1,2,\dots,n\}$. For $i,j\in [n]$, $i\leq j$, let $[i,j]=\{i,i+1,\dots,j\}$.
\item For any $m\times n$ matrix $M=(a_{i,j})_{i\in [m],j\in [n]}$, and $I\subset [m]$, $J\subset [n]$, let $M_{I}=(a_{i,j})_{i\in I,j\in [n]}$, $M^{J}=(a_{i,j})_{i\in [m],j\in J}$, $M_{I}^{J}=(a_{i,j})_{i\in I,j\in J}$.
\item For any $n$-tuple $\mathbf{v}=(v_1,\dots, v_n)$, denote by $\mathbf{v}[i]$ the $i$-th entry of $\mathbf{v}$, i.e., $\mathbf{v}[i]=v_i$ for $1\leq i\leq n$.
\item Given two integers $n\geq 2$ and $N \geq 0$, denote by $\langle N\rangle_n$ the remainder of the division of $N$ by $n$.

\end{itemize}

\subsection{Cyclic Codes and Constacyclic Codes}
Since cyclic codes can be seen as a special case of constacyclic codes, here we only introduce some basic results on constacyclic codes. The readers may refer to some standard textbooks on coding theory for more details (\cite{HP03}, \cite{MS77}).

An $[n,k]_q$ code is just a $k$-dimensional subspace of $\mathbb{F}_{q}^{n}$.
Let $\lambda\in \mathbb{F}_{q}^{*}=\mathbb{F}_{q}\setminus\{0\}$. We call the linear code $\mathcal{C}$ a $\lambda$-constacyclic code if $(\lambda c_{n-1},c_0,\dots,c_{n-2})\in \mathcal{C}$ provided any codeword $(c_0,c_1,\dots,c_{n-1})\in \mathcal{C}$. If we correspond a codeword $(c_0,c_1,\dots,c_{n-1})\in \mathcal{C}$ to a polynomial $\sum_{i=0}^{n-1}c_ix^{i}\in \mathbb{F}_{q}[x]$, then $\mathcal{C}$ can be regarded as an ideal of the quotient ring of $\mathbb{F}_{q}[x]/\langle x^n-\lambda\rangle$. Since $\mathbb{F}_{q}[x]/\langle x^n-\lambda\rangle$ is a principal ideal domain, $\mathcal{C}$ can be generated by a unique monic polynomial $g(x) \in \mathbb{F}_{q}[x]$ of the smallest degree such that $g(x)\mid (x^n-\lambda)$, and $g(x)$ is called the generator polynomial of $\mathcal{C}$.

Denote $\kappa$ the order of $\lambda$, let $s$ be the least integer such that $\kappa n\mid (q^s-1)$, then there exists a $\kappa n$-th root of unity $\eta\in \mathbb{F}_{q^{s}}$ such that $\eta^{n}=\lambda$. Let $\xi=\eta^{\kappa}$ be a primitive $n$-th root of unity, then $x^n-\lambda=\prod_{i=0}^{n-1}(x-\eta\xi^{i})$. When $\gcd(n,q)=1$, $g(x)$ has no repeated roots, the zeros of $g(x)$ form a subset of $\{\eta\xi^{i}\}_{i=0}^{n-1}$. Define $S_{\mathcal{C}}=\{0\leq i\leq n-1|~g(\eta\xi^{i})=0\}$. Since $\mathcal{C}$ is fully determined by $g(x)$, and $g(x)$ is fully determined by $S_{\mathcal{C}}$ whenever $\eta$ and $\xi$ are selected, we call $S_{\mathcal{C}}$ the defining set of $\mathcal{C}$.

When $\lambda=1$, $\lambda$-constacyclic codes degenerate to cyclic codes. By the following proposition, we can get lower bounds for the minimum distance of constacyclic codes from lower bounds for the minimum distance of cyclic codes.

\begin{proposition}\label{prop:boundforconsta}
Suppose that $\mathcal{C}$ is a cyclic $[n,k,d]$ code over $\mathbb{F}_{q^{m}}$ with generator polynomial $g_1(x)=\prod_{i\in S}(x-\xi^{i})$, where $m$ is the least integer such that $n\mid (q^m-1)$, $\xi \in \mathbb{F}_{q^m}$ is a primitive $n$-th root of unity, and $S\subset \{0,1,\dots,n-1\}$.
Let $\lambda\in \mathbb{F}_{q}^{*}$ and $\mathcal{C}'$ be a $\lambda$-constacyclic $[n,k',d']$ code over $\mathbb{F}_{q}$ with generator polynomial $g_2(x)$ whose zeros form a set containing $\{\eta\xi^{j}~|~j\in S\}$, where $\eta^n=\lambda$, then $d'\geq d$.
\end{proposition}

\begin{IEEEproof}
Assume $S=\{i_{1},i_{2},\dots,i_{u}\}\subset \{0,1,\dots,n-1\}.$
Then $\mathcal{C}$ has a parity check matrix
\[H=\begin{pmatrix}
1 & \xi^{i_{1}} & \xi^{2i_{1}} & \dots & \xi^{(n-1)i_{1}}\\
1 & \xi^{i_{2}} & \xi^{2i_{2}} & \dots & \xi^{(n-1)i_{2}}\\
\vdots & \vdots & \vdots & \ddots & \vdots\\
1 & \xi^{i_{u}} & \xi^{2i_{u}} & \dots & \xi^{(n-1)i_{u}}
\end{pmatrix}.\]
For any $J\subset [n]$ with $|J|=d-1$, $H^{J}$ is of full column rank. Thus we can find $I\subset [u]$ with $|I|=d-1$ such that ${\rm det}(H_{I}^{J})\neq 0$.

On the other hand, for any codeword $\mathbf{c}\in \mathcal{C}'$, we have $\widetilde{H}\mathbf{c}^{T}=\mathbf{0}$, where
\[\widetilde{H}=\begin{pmatrix}
1 & \eta\xi^{i_{1}} & \eta^{2}\xi^{2i_{1}} & \dots & \eta^{n-1}\xi^{(n-1)i_{1}}\\
1 & \eta\xi^{i_{2}} & \eta^{2}\xi^{2i_{2}} & \dots & \eta^{n-1}\xi^{(n-1)i_{2}}\\
\vdots & \vdots & \vdots & \ddots & \vdots\\
1 & \eta\xi^{i_{u}} & \eta^{2}\xi^{2i_{u}} & \dots & \eta^{n-1}\xi^{(n-1)i_{u}}
\end{pmatrix}.\]
It can be easily checked that $\widetilde{H}_{I}^{J}=H_{I}^{J}D$, where $D={\rm diag} [\eta^{t-1}]_{t\in J}$, hence ${\rm det}(\widetilde{H}_{I}^{J})\neq 0$. As $J\subset [n]$ is chosen arbitrarily, which means any $d-1$ columns of $\widetilde{H}$ are linearly independent. Assume that ${\rm wt}(\mathbf{c})=w<d$, from $\widetilde{H}\mathbf{c}^{T}=\mathbf{0}$, there exist $w$ columns in $\widetilde{H}$ which are linearly dependent, contradiction. Consequently, $d'\geq d$.
\end{IEEEproof}

\subsection{Cyclic and Constacyclic Locally Repairable Codes}

The following lemma provides a sufficient condition for a constacyclic or cyclic code to have $(r,\delta)$-locality.
\begin{lemma}\label{lem:locality}(See \cite[Lemma~2]{Sun19})
Suppose that $\gcd(n,q)=1$, $(r+\delta-1)\mid n$, and $\rho=\frac{n}{r+\delta-1}$. Let $\mathcal{C}$ be a $\lambda$-constacyclic code of length $n$ over $\mathbb{F}_{q}$ with generator matrix $g(x)\in \mathbb{F}_q[x]$. If there
exist integers $\varepsilon_0$ and $t > 0$ with $\gcd(t, r + \delta-1) = 1$ such
that $g(\eta\xi^{[\varepsilon_0+ti+j(r + \delta-1)]}) = 0$, for all $0 \leq i \leq \delta- 2$ and
$0 \leq j \leq \rho -1$, then $\mathcal{C}$ has $(r,\delta)$-locality.
\end{lemma}

\subsection{Singleton Procedure}

This subsection is devoted to introducing the singleton procedure. We primarily use the terminology from \cite{Betti06} and have included additional content to enhance its comprehensiveness and accessibility.

\begin{definition}
Let $\mathcal{C}$ be a linear code and $\mathbf{c}=(c_0,c_1,\dots,c_{n-1}) \in \mathcal{C}$, the Discrete Fourier Transform of $\mathbf{c}$ is defined by
    $${\rm DFT}(\mathbf{c})=(A_0,A_1,\dots,A_{n-1}), \textnormal{ where } A_{i}=\sum_{j=0}^{n-1}c_j\xi^{ij}.$$
\end{definition}

In \cite{Blahut95}, Blahut proved the following theorem, which builds a connection between the weight of a codeword and the rank of a matrix related to the codeword.

\begin{theorem}\label{thm:blahut}(Blahut's theorem, \cite{Blahut95})
If $\mathbf{c}=(c_0,c_1,\dots,c_{n-1})$ is a codeword of linear code $\mathcal{C}$, then
$${\rm wt}(\mathbf{c})={\rm rank}(M({\rm DFT}(\mathbf{c}))).$$

\end{theorem}

From Theorem~\ref{thm:blahut} we can see that the minimum distance of $\mathcal{C}$ is equal to the minimum rank of matrices in $\{M({\rm DFT}(\textbf{c})): \textbf{c}\in \mathcal{C}\setminus \{\mathbf{0}\}\}$.

\begin{theorem}\label{thm:schuab}(Schuab's theorem, \cite{Sch88})
If $\mathcal{C}$ is a linear code, then
$$d(\mathcal{C})=\min\{{\rm rank}(M({\rm DFT}(\mathbf{c})))~|~\mathbf{c}\in \mathcal{C}\setminus \{\bm 0\}\}.$$
\end{theorem}

Generally, it is difficult to determine the minimum distance directly from Theorem~\ref{thm:schuab} since all codewords need to be traversed. But for cyclic codes, the authors in \cite{Betti06} leveraged the technique of the singleton procedure proposed in \cite{Pon03} to establish a lower bound for ${\rm rank}(M({\rm DFT}(\mathbf{c})))$ and subsequently presented a lower bound for the minimum distance of the cyclic codes.

 The relation of the DFT of a codeword in a cyclic code with the defining set is given by the following fact.

\textbf{Fact 1}: Suppose $\mathcal{C}$ is a cyclic code. Let $\mathbf{c}=(c_0,c_1,\dots,c_{n-1})$ be a codeword of $\mathcal{C}$ and ${\rm DFT}(\mathbf{c})=(A_0,A_1,\dots,A_{n-1})$ be the DFT of $\mathbf{c}$. Then $A_i=0$ for $i\in S_{\mathcal{C}}$.

The fact immediately follows by $A_{i}=c(\xi^{i})$, where $c(x)=\sum_{i=0}^{n-1}c_ix^{i}$.

Before introducing the singleton procedure, we first provide several related notions and definitions.
\begin{definition}\label{def:U}
Let $\mathbb{K}$ be a field and $\mathbf{v}=(v_1,\dots,v_{n})\in \mathbb{K}^{n}$, for each $v_i$, $1\leq i\leq n$, we say that
    \begin{itemize}
        \item $v_{i}$ is $\Delta^{+}$, if we know for sure that $v_{i}$ is different from $0$;
        \item $v_{i}$ is $0$, if we know for sure that $v_{i}$ is $0$;
        \item $v_{i}$ is $\Delta$, otherwise.
    \end{itemize}
Denote $\mathcal{U}=\{\Delta,\Delta^{+},0\}$ and $\mathcal{U}^{n}$ the set of all $n$-tuples whose components are elements in $\mathcal{U}$.
For any vector $\mathbf{v}$ in $\mathbb{K}^{n}$, we can represent $\mathbf{v}$ by an $n$-tuple $\mathbf{u}$ in $\mathcal{U}^{n}$ based on the available information. We term $\mathbf{u}$ as an $\mathcal{U}$-\textbf{representation} of $\mathbf{v}$, and denote it as $\mathbf{v}_{\mathcal{U}}$.
\end{definition}

One should regard an $\mathcal{U}$-representation of $\mathbf{v}$ as the information we have on $\mathbf{v}$, rather than a way to indicate its value. For cyclic codes, all the information about their codewords' DFTs derives from their definition sets.

\begin{definition}
 Let $\mathcal{C}$ be a cyclic code of length $n$ with defining set
$S_{\mathcal{C}}$. We denote by $R(n, S_{\mathcal{C}})$ the vector $(u_1,\dots,u_n) \in \mathcal{U}^n$ such that
\[u_i=\left\{
\begin{array}{lcl}
0,& &  {\rm if }\,\, i-1\,\in\, S_{\mathcal{C}}, \\
\Delta, & & {\rm otherwise }.
 \end{array} \right. \]

\end{definition}

\begin{definition}\label{def:04}
Given a vector $\textbf{v} \in \mathcal{U}^n$, we denote by $\mathcal{A}(\textbf{v})$ the set
of vectors $\mathbf{u} \in \mathcal{U}^n\setminus \{\bm 0\}$ such that
\begin{itemize}
  \item $\mathbf{u}[i] = 0$, if $\mathbf{v}[i] = 0$;
  \item $\mathbf{u}[i] = \Delta^{+}$, if $\mathbf{v}[i] =\Delta^{+}$;
  \item $\mathbf{u}[i] = \Delta^{+}$, or $\mathbf{u}[i] = 0$ if $\mathbf{v}[i] =\Delta$.
\end{itemize}
\end{definition}
Observe that if there is at least one component of $\textbf{v}$ equal to $\Delta^{+}$ then
$|\mathcal{A}(\textbf{v})| = 2^s$, where $s$ represents the number of components of $\textbf{v}$ equal
to $\Delta$. Otherwise $|\mathcal{A}(\textbf{v})| = 2^s-1$.

From \textbf{Fact 1} and Definitions~\ref{def:U}-\ref{def:04}, we can get the following lemma directly.
\begin{lemma}\label{lem:abstraction}
Let $\mathcal{C}$ be a cyclic code of length $n$ with defining set
$S_{\mathcal{C}}$, then \[\{{\rm DFT}(\mathbf{c}): \mathbf{c}\in \mathcal{C}\setminus\{\mathbf{0}\}\} \subseteq \mathcal{A}(R(n,S_{\mathcal{C}})).\]
\end{lemma}

Now, we present the formal description of the singleton procedure.

\begin{definition}
Let $M$ be a matrix over $\mathcal{U}$. We say that column $M^{\{j\}}$ is a \textbf{singleton} if it has only one nonzero component $M^{\{j\}}_{\{i\}}$, i.e., $M^{\{j\}}_{\{i\}} = \Delta^{+}$ and $M^{\{j\}}_{\{\ell\}} = 0$ for $\ell \neq i$. When this happens, we say that the $i$-th row is the row corresponding to the singleton.

A set of $r$ row vectors of length $n$, with $r \leq n$, form a matrix
$A_r \in \mathcal{U}^{r\times n}$. If a column $A_r^{\{j\}}$ is a singleton, we can erase from $A_r$
the $j$-th column and the corresponding row, we refer to this operation as a \textbf{singleton-deletion} related to the singleton. We denote by $A_{r-1}$ the $(r-1)\times(n-1)$ matrix obtained after a singleton-deletion and search for a new singleton in $A_{r-1}$. A \textbf{singleton procedure} is nothing but a procedure repeating searching for a new singleton and implementing the related singleton-deletion. If this procedure can continue until we have obtained a $1\times (n-r +1)$ matrix $A_1$ containing at least one $\Delta^{+}$, then we say that the singleton procedure is \textbf{successful} for the set of $r$ rows.
\end{definition}

\begin{example}\label{exa:00}
 For the upper-triangle or lower-triangle matrices, there is a trivial successful singleton procedure for all their rows. We indicate the order of singleton-deletions beside the matrices in the following examples.
\[   \begin{array}{r}
  {\rm 1st~ step} \\
  {\rm 2nd~ step}\\
  {\rm 3rd~ step}\\
  {\rm 4th~ step}\\
  {\rm 5th~ step}
     \end{array}
 \left[\begin{array}{ccccc}
    \Delta^{+}&\Delta&\Delta&\Delta&\Delta \\
    0&\Delta^{+}&\Delta&\Delta&\Delta \\
    0&0&\Delta^{+}&\Delta&\Delta \\
    0&0&0&\Delta^{+}&\Delta \\
    0&0&0&0&\Delta^{+} \\
  \end{array}\right]
\]
\[
  \begin{array}{l}
  {\rm 5th~ step} \\
  {\rm 4th~ step}\\
  {\rm 3rd~ step}\\
  {\rm 2nd~ step}\\
  {\rm 1st~ step}
     \end{array}
  \left[\begin{array}{ccccc}
    \Delta^{+}&0&0&0&0\\
    \Delta&\Delta^{+}&0&0&0\\
    \Delta&\Delta&\Delta^{+}&0&0\\
    \Delta&\Delta&\Delta&\Delta^{+}&0\\
    \Delta&\Delta&\Delta&\Delta&\Delta^{+}
  \end{array}\right]
\]
\end{example}

\begin{definition}
Given a matrix $M$ over $\mathcal{U}$, we denote by ${\rm prk}(M)$
the pseudo-rank of $M$, i.e., the largest $r$ such that there exists a set of
$r$ rows in $M$ for which the singleton procedure is successful.
\end{definition}

The following lemma builds a connection between rank and pseudo-rank.
\begin{lemma}\label{lem:boundforrank}
For any codeword $\mathbf{c}$ in a cyclic code $\mathcal{C}$, denote by $\mathbf{v}$ the {\rm DFT} of $\mathbf{c}$, then
$${\rm rank}(M(\mathbf{v}))\geq {\rm prk}(M(\mathbf{v}_{\mathcal{U}})).$$
\end{lemma}
\begin{IEEEproof}
It is evident that the $i$-th row of $M(\mathbf{v}_{\mathcal{U}})$ is the $\mathcal{U}$-representation of the
 $i$-th row of $M(\mathbf{v})$. And the $\Delta^{+}$'s and $0$'s in $M(\mathbf{v}_{\mathcal{U}})$ correspond to nonzero elements and zeros in $M(\mathbf{v})$ respectively at the same positions.

Assume ${\rm prk}(M(\mathbf{v}_{\mathcal{U}}))=s$, which means that there is a successful singleton procedure for $r$ rows of $M(\mathbf{v}_{\mathcal{U}})$. Let $\overline{\mathbf{r}}_i$ be the row corresponding to the $i$-th singleton after the first $i-1$ singleton-deletions in $M(\mathbf{v}_{\mathcal{U}})$, and $\mathbf{r}_i$
be the row in $M(\mathbf{v})$ which has the same row index as $\overline{\mathbf{r}}_i$ has in $M(\mathbf{v}_{\mathcal{U}})$, for all $i\in [s]$.

It can be easily checked that, for $i\in [s]$, there exists a position where $\mathbf{r}_i$ has a nonzero element and all the $\mathbf{r}_j$'s with $i+1\leq j\leq s$ have zeros. Hence $\mathbf{r}_1$ is linearly independent to $\{\mathbf{r}_2,\dots, \mathbf{r}_s\}$, $\mathbf{r}_2$ is linearly independent to $\{\mathbf{r}_3,\dots \mathbf{r}_s\}$, $ \dots$, $\mathbf{r}_{s-1}$ is linearly independent to $\mathbf{r}_s$. Subsequently $\mathbf{r}_1,\dots,\mathbf{r}_s$ are linearly independent, i.e., ${\rm rank}(M(\mathbf{v}))\geq s$.
The proof is completed.
\end{IEEEproof}

Building upon Lemmas~\ref{lem:abstraction} and \ref{lem:boundforrank}, as well as Schuab's theorem, the following theorem can be deduced.

\begin{theorem}\label{thm:Betti}(See \cite[Theorem~2.1]{Betti06})
Let $\mathcal{C}$ be a cyclic code of length $n$, defining
set $S_{\mathcal{C}}$ and minimum distance $d$. Then
$$d \geq \min\{{\rm prk}(M(\mathbf{u}))~|~\mathbf{u}\in \mathcal{A}(R(n,S_{\mathcal{C}}))\}.$$
\end{theorem}

Here are some lemmas that will be helpful in the proofs of our theorems.
\begin{lemma}\label{lem:shift}(See \cite[Lemma~3.1]{Betti06})
Let $\mathbf{u}, \mathbf{v} \in \mathcal{U}^{n}$ and $m \in \mathbb{N}$. If $\mathbf{u}$ is obtained by a cyclic shift of $\mathbf{v}$ by $m$ places then ${\rm prk}(M(\mathbf{u})) = {\rm prk}(M(\mathbf{v}))$.
\end{lemma}

\begin{lemma}\label{lem:bch}(See \cite[Lemma~3.2]{Betti06})
If $\mathbf{v} \in \mathcal{U}^{n}$ has the form
$$(\overbrace{0,\dots,0}^{r},\Delta^{+},\dots)$$
and $r\geq 0$, then ${\rm prk}(M(\mathbf{v}))\geq r+1$.
\end{lemma}

By utilizing the singleton procedure with the above theorems and lemmas, Betti and Sala proposed the following bound.

\begin{theorem}\label{thm:B-S}\cite[Theorem~3.1]{Betti06} (Betti-Sala bound)
Let $\mathcal{C}$ be a cyclic code of length $n$ and defining
set $S_{\mathcal{C}}$.
Suppose that there are $\delta,s \in \mathbb{N}$ with $\delta\geq 1$, $s\geq 1$ and $i_0\in \{0,1,\dots,n-1\}$ such that:

\begin{itemize}
  \item[(1)] $\langle i_0+j\rangle_{n}\in S_{\mathcal{C}}$, for any $0\leq j\leq s\delta-1$,
  \item[(2)] $\langle i_0+j\rangle_{n}\in S_{\mathcal{C}}$, for any $j=(s+h)\delta+1,\dots,(s+h)\delta+\delta-1$, $0\leq h\leq s$.
\end{itemize}
Then
$$d(\mathcal{C})\geq (s+1)\delta.$$
\end{theorem}

From the assumption of the Theorem~\ref{thm:B-S}, $R(n, S_{\mathcal{C}})$ contains (allowing for wrapping) a consecutive part of
\begin{equation}\label{equa:00}
\overbrace{0,\dots,0}^{s\delta},\overbrace{\underline{\Delta,\overbrace{0,\dots,0}^{\delta-1}},\dots,\underline{\Delta,\overbrace{0,\dots,0}^{\delta-1}}}^{(s+1)\delta}.
\end{equation}
For any sequence $\mathcal{S}$ with components in $\mathcal{U}$, we denote $(\mathcal{S})^m$ the $m$-repetition of $\mathcal{S}$ for any $m\in \mathbb{Z}^{+}$. For example, $(0,\Delta)^{2}=(0,\Delta,0,\Delta)$.
And we can rewrite \eqref{equa:00} as
$$(0)^{s\delta}((\Delta)^{1}(0)^{\delta-1})^{s+1}.$$

\section{New Bounds from Singleton Procedure Technique}

In this section, we present our new bounds for the minimum distance of cyclic codes via the singleton procedure technique.

\subsection{On The Pseudo-Rank of a Special Class of Square Matrices}
In this subsection, we provide a criterion for determining the pseudo-rank of a square matrix with a special type, which serves as a foundation for our derivation of bounds for the minimum distance of cyclic codes in the subsequent subsections.

For any $\mathbf{x}=(x_1,x_2,\dots,x_n) \in \mathcal{U}^{n}$, let ${\rm zero}(\mathbf{x})=\{i~|~x_{i}=0 \textnormal{ and } 1\leq i\leq n\}$, $\mathbf{x}^{-1}=(x_n,x_{n-1},\dots,x_1)$.
For any two $n$-tuples $\mathbf{u}=(u_1,u_2,\dots,u_n),\mathbf{v}=(v_1,v_2,\dots,v_n)$ in $\mathcal{U}^{n}$,
we denote $\mathbf{u} \subseteq_{0} \mathbf{v}$ if ${\rm zero}(\mathbf{u})\subseteq {\rm zero}(\mathbf{v})$.

\begin{lemma}\label{lem:main}
Let $A$ be an $m\times m$ matrix over $\mathcal{U}$ with the following properties:
\begin{description}
  \item[(a)] $A_{\{i\}}^{\{j\}}=A_{\{i+1\}}^{\{j+1\}}$ for $1\leq i, j\leq m-1$,
  \item[(b)] $A^{\{1\}}=((\Delta^{+})^{1}(0)^{m-1-t_0}(\Delta)^{t_0})^{T}$, where $t_0\in \mathbb{N}^{+}$ (when $t_0=0$, $A$ is an upper-triangle matrix) and
  \item[(c)] $A^{\{m\}}=((0)^{z_1} (\Delta)^{t_1}(0)^{z_2}(\Delta)^{t_2}\dots(0)^{z_s}(\Delta)^{t_s}(0)^{z_{s+1}}(\Delta^{+})^{1})^{T},$ where $s\geq 1$, $t_i\in \mathbb{N}^{+}$ for $1\leq i\leq s$, $z_i\in \mathbb{N}^{+}$ for $1\leq i\leq s+1$ and $\sum_{i=1}^{s}(t_i+s_i)+z_{s+1}+1=m$.
\end{description}

Let $x_i:=z_i+t_i$, $y_i:=\sum_{j=1}^{i}x_j$ and
\[
\mathcal{S}_{i}=((\Delta^{+})^{1}(0)^{x_i-1}(\Delta)^{x_{i+1}-(t_{i+1}+z_i-1)}(0)^{t_{i+1}+z_i-1}\dots(\Delta)^{x_{s}-(t_s+z_i-1)}(0)^{t_s+z_i-1}(\Delta)^{m-y_{s}+y_{i-1}})\] for $1\leq i\leq s$, where we set $y_0=0$. If

\begin{itemize}
  \item[(1)] $t_0\leq z_1$ and
  \item[(2)] $\mathcal{S}_{i} \subseteq_{0} ((A^{\{m\}})^{T})^{-1}$ for all $1\leq i\leq s$,
\end{itemize}
then ${\rm prk}(A)= m$.
\end{lemma}

The proof of Lemma~\ref{lem:main} is quite technical, and we would prefer to illustrate it with a preceding example.

\begin{example}\label{example:01}
Let $s=2$, we have
\[
\begin{aligned}
((A^{\{m\}})^{T})^{-1} &=\left((\Delta^{+})^{1}(0)^{z_{3}}(\Delta)^{t_2}(0)^{z_2}(\Delta)^{t_1}(0)^{z_1}\right),\\
\mathcal{S}_{1}&=\left((\Delta^{+})^{1}(0)^{z_1+t_1-1}(\Delta)^{z_2-z_1+1}(0)^{t_{2}+z_1-1}(\Delta)^{m-y_{2}}\right),\\ \mathcal{S}_{2}&=\left((\Delta^{+})^{1}(0)^{z_2+t_2-1}(\Delta)^{m-y_{2}+y_1}\right).
\end{aligned}
\]
Then $\mathcal{S}_{1} \subseteq_{0} ((A^{\{m\}})^{T})^{-1}$ is equivalent to the following inequalities

$$
\left\{
\begin{aligned}
z_1+t_1&\leq z_3+1,\\
z_1+t_1+z_2-z_1+1&\geq 1+z_{3}+t_2,\\
z_1+t_1+z_2+t_2 &\leq 1+z_{3}+t_2+z_2,
\end{aligned}
\right.
$$
and $\mathcal{S}_{2} \subseteq_{0} ((A^{\{m\}})^{T})^{-1}$ is equivalent to the following inequality
$$z_2+t_2\leq z_3+1.$$
It can be easily checked that $z_1=3$, $t_1=2$, $z_2=3$, $t_2=1$ $z_3=4$, $t_0=z_1=3$
satisfies the conditions (1) and (2) in Lemma~\ref{lem:main}.
The matrix $A$ in Lemma~\ref{lem:main} has the following form:

\[
\begin{tikzpicture}[scale=0.1]
 \matrix (m)[
    matrix of math nodes,
    nodes in empty cells,
    minimum width=width("6"),
  ] {
  1 & 2 & 3 & 4 & 5 & 6 & 7 & 8 & 9 & 10 & 11 & 12 & 13 & 14 \\
   \Delta^{+} & 0 & 0 & 0 & 0 & \Delta & 0 & 0 & 0 & \Delta & \Delta & 0 & 0 & 0 & 1 \\
 0& \Delta^{+}  & 0 & 0 & 0 & 0 & \Delta & 0 & 0 & 0 & \Delta & \Delta & 0 & 0 & 2 \\
 0&  0& \Delta^{+}  & 0 & 0 & 0 & 0 & \Delta & 0 & 0 & 0 & \Delta & \Delta & 0 & 3 \\
 0& 0&  0& \Delta^{+}  & 0 & 0 & 0 & 0 & \Delta & 0 & 0 & 0 & \Delta & \Delta & 4 \\
 0 &0& 0&  0& \Delta^{+}  & 0 & 0 & 0 & 0 & \Delta & 0 & 0 & 0 & \Delta & 5 \\
 0& 0 &0& 0&  0& \Delta^{+}  & 0 & 0 & 0 & 0 & \Delta & 0 & 0 & 0 & 6 \\
 0& 0& 0 &0& 0&  0& \Delta^{+}  & 0 & 0 & 0 & 0 & \Delta & 0 & 0 & 7 \\
 0& 0& 0& 0 &0& 0&  0& \Delta^{+}  & 0 & 0 & 0 & 0 & \Delta & 0 & 8 \\
 0& 0& 0& 0& 0 &0& 0&  0& \Delta^{+}  & 0 & 0 & 0 & 0 & \Delta & 9 \\
 0& 0& 0& 0& 0& 0 &0& 0&  0& \Delta^{+}  & 0 & 0 & 0 & 0 & 10 \\
 0& 0& 0& 0& 0& 0& 0 &0& 0&  0& \Delta^{+}  & 0 & 0 & 0 & 11 \\
 \Delta& 0& 0& 0& 0& 0& 0& 0 &0& 0&  0& \Delta^{+}  & 0 & 0 & 12\\
 \Delta& \Delta& 0& 0& 0& 0& 0& 0& 0 &0& 0&  0& \Delta^{+}  & 0 & 13\\
 \Delta& \Delta& \Delta& 0& 0& 0& 0& 0& 0& 0 &0& 0&  0& \Delta^{+}  & 14\\
  } ;
\draw (m-4-4.south west) rectangle (m-2-5.north east);
\draw (m-4-9.south west) rectangle (m-2-9.north east);
\draw (m-9-9.south west) rectangle (m-7-9.north east);
\end{tikzpicture}.
\]

Our goal is to prove that there is a successful singleton procedure for all the rows of $A$.
Observe that if the last $t_0=3$ rows of $A$ can be eliminated after a singleton procedure for \textbf{some} rows of $A$, the remaining submatrix is an upper-triangle matrix, which has a trivial successful singleton procedure. Then we will have a successful singleton procedure for \textbf{all} the rows of $A$.

Hence it is only need to prove that the last three rows of $A$ can be eliminated by some singleton-deletions. The goal can be achieved in the following manner.

(1) Eliminate the last row of $A$.

\begin{itemize}
 \item Let $A_{1,1}=A_{\{1,2,3\}}^{\{4,5\}}$, $A_{1,2}=A_{\{1,2,3\}}^{\{9\}}$, $A_{2,2}=A_{\{6,7,8\}}^{\{9\}}$ (blocked parts in $A$), as we can see, these three submatrices are all-zero matrices.
  \item As $3=t_0\leq z_1=3$ and $A_{1,1}$ is an all-zero matrix, the columns in which $A_{1,1}$ is located, i.e., $A^{\{4\}}, A^{\{5\}}$, are singletons. We use these columns to perform singleton-deletions.
  \item Since $A_{1,2}$ and $A_{2,2}$ are all-zero matrices, the column in which $A_{2,2}$ is located, i.e., $A^{\{9\}}$ becomes a singleton. We use this column to perform singleton-deletion.
 \item The last column, i.e., $A^{\{14\}}$ becomes a singleton, as all the $\Delta$'s in $A^{\{14\}}$ were eliminated. We use $A^{\{14\}}$ to perform singleton-deletion, the last row of $A$ is eliminated.
\end{itemize}

(2) Eliminate the second last row of $A$.

\begin{itemize}
 \item After the aforementioned singleton-deletions, matrix $A$ becomes:

\[
\begin{tikzpicture}[scale=0.5]
 \matrix (m)[
    matrix of math nodes,
    nodes in empty cells,
    minimum width=width("6"),
  ] {
  1 & 2 & 3 & 4 & 5 & 6 & 7 & 8 & 9 & 10 & 11 & 12 & 13 & 14 \\
  \Delta^{+} & 0 & 0 & \quad & \quad  & \Delta & 0 & 0 & \quad & \Delta & \Delta & 0 & 0 &  & 1 \\
 0& \Delta^{+}  & 0 &  &  & 0 & \Delta & 0 &  & 0 & \Delta & \Delta & 0 &  & 2 \\
 0&  0& \Delta^{+}  &  &  & 0 & 0 & \Delta &  & 0 & 0 & \Delta & \Delta &  & 3 \\
 & &  &   &  &  &  &  &  &  &  &  &  &  & 4 \\
 & &  &   &  &  &  &  &  &  &  &  &  &  & 5 \\
 0& 0 &0& &  & \Delta^{+}  & 0 & 0 &  & 0 & \Delta & 0 & 0 &  & 6 \\
 0& 0& 0 && &  0& \Delta^{+}  & 0 &  & 0 & 0 & \Delta & 0 &  & 7 \\
 0& 0& 0&  && 0&  0& \Delta^{+}  &  & 0 & 0 & 0 & \Delta &  & 8 \\
 & &  &   &  &  &  &  &  &  &  &  &  &  & 9 \\
 0& 0& 0& & & 0 &0& 0&  & \Delta^{+}  & 0 & 0 & 0 &  & 10 \\
 0& 0& 0& & & 0& 0 &0& &  0& \Delta^{+}  & 0 & 0 &  & 11 \\
 \Delta& 0& 0& & & 0& 0& 0 && 0&  0& \Delta^{+}  & 0 &  & 12\\
 \Delta& \Delta& 0& & & 0& 0& 0&  &0& 0&  0& \Delta^{+}  &  & 13\\
 & &  &   &  &  &  &  &  &  &  &  &  &  &  14\\
  } ;
\draw (m-3-3.south west) rectangle (m-2-3.north east);
\draw (m-8-8.south west) rectangle (m-7-8.north east);
\draw (m-3-8.south west) rectangle (m-2-8.north east);
\end{tikzpicture}.
\]
  \item The column $A^{\{3\}}$ becomes a singleton. We use this column to perform singleton-deletion.
  \item The column $A^{\{8\}}$ becomes a singleton. We use this column to perform singleton-deletion.
  \item The column $A^{\{13\}}$ becomes a singleton, as all the $\Delta$'s in $A^{\{13\}}$ were eliminated. We use $A^{\{13\}}$ to perform singleton-deletion, the second last of the row of $A$ is eliminated.
\end{itemize}

(3)  Eliminate the third last row of $A$.

\begin{itemize}
 \item After the aforementioned singleton-deletions, matrix $A$ becomes:

\[
\begin{tikzpicture}[scale=0.5]
 \matrix (m)[
    matrix of math nodes,
    nodes in empty cells,
    minimum width=width("6"),
  ] {
  1 & 2 & 3 & 4 & 5 & 6 & 7 & 8 & 9 & 10 & 11 & 12 & 13 & 14 \\
  \Delta^{+} & 0 & \quad & \quad & \quad  & \Delta & 0 & \quad & \quad & \Delta & \Delta & 0 & \quad & \quad & 1 \\
 0& \Delta^{+}  &  &  &  & 0 & \Delta &  &  & 0 & \Delta & \Delta &  &  & 2 \\
 & &  &   &  &  &  &  &  &  &  &  &  &  & 3 \\
 & &  &   &  &  &  &  &  &  &  &  &  &  & 4 \\
 & &  &   &  &  &  &  &  &  &  &  &  &  & 5 \\
 0& 0 && &  & \Delta^{+}  & 0 &  &  & 0 & \Delta & 0 &  &  & 6 \\
 0& 0&  && &  0& \Delta^{+}  &  &  & 0 & 0 & \Delta &  &  & 7 \\
 & &  &   &  &  &  &  &  &  &  &  &  &  & 8 \\
 & &  &   &  &  &  &  &  &  &  &  &  &  & 9 \\
 0& 0& & & & 0 &0& &  & \Delta^{+}  & 0 & 0 &  &  & 10 \\
 0& 0& & & & 0& 0 && &  0& \Delta^{+}  & 0 & &  & 11 \\
 \Delta& 0& & & & 0& 0&  && 0&  0& \Delta^{+}  &  &  & 12\\
 & &  &   &  &  &  &  &  &  &  &  &  &  & 13\\
 & &  &   &  &  &  &  &  &  &  &  &  &  &  14\\
  } ;
\draw (m-2-2.south west) rectangle (m-2-2.north east);
\draw (m-2-7.south west) rectangle (m-2-7.north east);
\draw (m-7-7.south west) rectangle (m-7-7.north east);

\end{tikzpicture}.
\]
  \item The column $A^{\{2\}}$ becomes a singleton. We use this column to perform singleton-deletion.
  \item The column $A^{\{7\}}$ becomes a singleton. We use this column to perform singleton-deletion.
 \item $A^{\{12\}}$ becomes a singleton, as all the $\Delta$'s in $A^{\{12\}}$ were eliminated. We use $A^{\{12\}}$ to perform singleton-deletion, the third last of the row of $A$ is eliminated.
\end{itemize}

\end{example}

\begin{remark}
In fact, $\mathcal{S}_{1} \subseteq_{0} ((A^{\{m\}})^{T})^{-1}$ is essentially a condition that ensures both $A_{1,1}$ and $A_{1,2}$ are all-zero matrices, while $\mathcal{S}_{2} \subseteq_{0} ((A^{\{m\}})^{T})^{-1}$ is a condition that ensures $A_{2,2}$ is an all-zero matrix. The contribution of $t_0\leq z_1$ is two-fold.
First, it serves as a necessary condition for the columns of $A$ containing $A_{1,1}$ to be natural singletons. Second, it ensures that the last $t_0$ rows of $A$ can always be eliminated.
\end{remark}

\begin{remark}\label{rem:deletion}
From the proof of Example~\ref{example:01}, we can observe that to proceed with a singleton procedure for certain rows of matrix $A$, it is necessary to eliminate some $\Delta$'s in $A$ by singleton-deletions. Here we provide a key observation that assists us in implementing singleton-deletions to eliminate targeted $\Delta$'s.
In the matrix $A$, $\Delta^{+}$'s only lie on the main diagonal. So in a singleton procedure for rows of $A$, singletons always have their only $\Delta^{+}$ in the positions of $A_{\{i\}}^{\{i\}}$ for some $i\in [m]$. To eliminate some $\Delta$ in position $A_{\{j\}}^{\{k\}}$, where $j\leq k\leq m$, it is necessary for the column $A^{\{j\}}$ to become a singleton.
\end{remark}

Now we can start the formal proof of Lemma~\ref{lem:main}.

\textit{Proof of Lemma~\ref{lem:main}:}
From the assumptions of the lemma, we can represent matrix $A$ using the following block matrix.

\begin{centering}
\resizebox{0.65\textwidth}{!}{%
  \begin{minipage}{\textwidth}
    \[
\begin{matrix}
  \begin{array}{|ccc|ccc|ccc|ccc|c|ccc|ccc|cccccc|}
\hline
\Delta^{+}&&&&&&&&&&&&&&&&&&&&&&&&0
\\[0.5em]
&\ddots&&&A_{1,1}&&&\ddots&&&A_{1,2}&&\cdots&&\ddots&&&A_{1,s}&&&&&&&\vdots \\[0.5em]
&&\Delta^{+}&&&&&&\blacktriangle&&&&&&&\blacktriangle&&&&&&&&\Delta&0\\[0.5em] \hline
&&&\Delta^{+}&&&&&&&&&&&&&&&&&&&&&\Delta\\[0.5em]
&&&&\ddots&&&*&&&*&&&&*&&&*&&&&&&&\vdots\\[0.5em]
&&&&&\Delta^{+}&&&&&&&&&&&&&&&&&& &\Delta \\[0.5em] \hline
&&&&&&\Delta^{+}&&&&&&&&&&&&&&&&&&0 \\[0.5em]
&&&&&&&\ddots&&&A_{2,2}&&\cdots&&\ddots&&&A_{2,s}&&&&&&&\vdots \\[0.5em]
&&&&&&&&\Delta^{+}&&&&&&&\blacktriangle&&&&&&&&\Delta&0 \\[0.5em] \hline
&&&&&&&&&\Delta^{+}&&&&&&&&&&&&&&&\Delta \\[0.5em]
&&&&&&&&&&\ddots&&&&*&&&*&&&&&&&\vdots\\[0.5em]
&&&&&&&&&&&\Delta^{+}&&&&&&&&&&&&&\Delta\\[0.5em] \hline
&&&&&&&&&&&&\ddots&&&\vdots&&\vdots&&&&&&\vdots&\vdots \\[0.5em] \hline
&&&&&&&&&&&&&\Delta^{+}&&&&&&&&&&&0\\[0.5em]
&&&&&&&&&&&&&&\ddots&&&A_{s,s}&&&&&&&\vdots\\[0.5em]
&&&&&&&&&&&&&&&\Delta^{+}&&&&&&&&\Delta&0\\[0.5em] \hline
&&&&&&&&&&&&&&&&\Delta^{+}&&&&&&&&\Delta\\[0.5em]
&&&&&&&&&&&&&&&&&\ddots&&&&&&&\vdots\\[0.5em]
&&&&&&&&&&&&&&&&&&\Delta^{+}&&&&&&\Delta\\[0.5em] \hline
&&&&&&&&&&&&&&&&&&&\Delta^{+}&&&&&0\\[0.5em]
&&&&&&&&&&&&&&&&&&&&\ddots&&&&\vdots\\[0.5em]
&&&&&&&&&&&&&&&&&&&&&\Delta^{+}&&&0\\[0.5em]
&L&&&&&&&&&&&&&&&&&&&&&\ddots&&\vdots\\[0.5em]
&&&&&&&&&&&&&&&&&&&&&&&\Delta^{+}&0\\[0.5em]
&&&&&&&&&&&&&&&&&&&&&&&&\Delta^{+} \\[0.5em] \hline
\end{array}
 \begin{aligned}
  &\left.\vphantom{\begin{array}{r}
  \partialphantom \\[0.33em]
  \partialphantom \\[0.33em]
  \partialphantom \\[0.33em]
  \end{array}} \right\} %
  z_1\\
  &\left.\vphantom{\begin{array}{r}
  \partialphantom \\[0.33em]
  \partialphantom \\[0.33em]
  \partialphantom \\[0.33em]
  \end{array}} \right\} %
   t_1\\
  &\left.\vphantom{\begin{array}{r}
  \partialphantom \\[0.33em]
  \partialphantom \\[0.33em]
  \partialphantom \\[0.33em]
  \end{array}} \right\} %
   z_2\\
  &\left.\vphantom{\begin{array}{r}
  \partialphantom \\[0.33em]
  \partialphantom \\[0.33em]
  \partialphantom \\[0.33em]
  \end{array}} \right\} %
   t_2\\
  &\begin{array}{r}
   \partialphantom  \\[0.5em]
  \end{array}\\ %
  &\left.\vphantom{\begin{array}{r}
  \partialphantom \\[0.33em]
  \partialphantom \\[0.33em]
  \partialphantom \\[0.33em]
  \end{array}} \right\} %
   z_{s}\\
  &\left.\vphantom{\begin{array}{r}
  \partialphantom \\[0.33em]
  \partialphantom \\[0.33em]
  \partialphantom \\[0.33em]
  \end{array}} \right\} %
   t_{s}\\
  &\left.\vphantom{\begin{array}{r}
  \partialphantom \\[0.33em]
  \partialphantom \\[0.33em]
  \partialphantom \\[0.33em]
  \partialphantom \\[0.33em]
  \partialphantom \\[0.33em]
  \end{array}} \right\} %
   z_{s+1}\\
  &\left.\vphantom{\begin{array}{r}
  \partialphantom \\[0.33em]
  \end{array}} \right\} %
   1\\
 \end{aligned}
\end{matrix}
    \]
  \end{minipage}
}
\end{centering}\\

where $A_{i,j}=A_{[y_{i-1}+1,y_{i-1}+z_i]}^{[y_j-t_j+1, y_j]}\in  \mathcal{U}^{z_{i}\times t_{j}}$ for $1\leq i\leq s$, $i\leq j\leq s$, and

\[
L=\begin{tabular}{r}
$\lefteqn{\phantom{\begin{matrix}  \vdots\\ a_0\ \end{matrix}}}$\\
$t_0\left\{ \lefteqn{\phantom{\begin{matrix}    b_0\\ \ddots \\ b_0\ \end{matrix}}} \right.$
\end{tabular} \hspace{-1em}
\left[\phantom{\begin{matrix} a_1\\a_0\\b_0\\ \ddots\\b_0 \end{matrix}}
\right.\hspace{-1em}
\underbrace{\begin{matrix}
\vdots& & & \\
0&  &  &  \\
\Delta & \ddots  &  & \\
\vdots & \ddots & \ddots & \\
\Delta & \cdots & \Delta & 0 & \cdots& \cdots
\end{matrix}}_{z_1}
\hspace{-1em}
\left.\phantom{\begin{matrix} \ddots\\a_0\\b_0\\ \ddots\\b_0 \end{matrix}}\right]\hspace{-1em}
\begin{tabular}{l}
$\left.\lefteqn{\phantom{\begin{matrix}a_0\\ c_1\\ b_0\\ \ddots\\ b_0\ \end{matrix}}} \right\}z_{s+1}+1$
\end{tabular}.
\]
From $A_{\{i\}}^{\{j\}}=A_{\{i+1\}}^{\{j+1\}}$ for $1\leq i, j\leq m-1$ and $t_0\leq z_1$, we know that $L$ is the only nonzero submatrix in the part below the main diagonal of $A$, i.e., the blank squares in $A$ represent all-zero matrices.

We use $\blacktriangle$'s to highlight the entries of $A_{\{y_{i-1}+z_i\}}^{\{y_{j-1}+z_j\}}$, where $1\leq i\leq s-1$, $i<j\leq s$, as they are crucial for our discussion.

As $A$ is an $m\times m$ square matrix, to prove ${\rm prk}(A)= m$, it is sufficient to find a successful singleton procedure for all the rows of $A$.
We then finish the proof by proving the following two claims.

\textbf{Claim 1}: If $A_{i,j}$ for $1\leq i\leq s$, $i\leq j\leq s$ are all-zero matrices, then there is a successful singleton procedure for all the rows of $A$.

We first do the following singleton-deletions to eliminate all the $\Delta$'s in the last column.
\begin{framed}

(1) As $A_{1,1}$ is an all-zero matrix, the columns in which $A_{1,1}$ is located are singletons. We use these columns to perform singleton-deletions.

(2) Since $A_{1,2}$ and $A_{2,2}$ are all-zero matrices, after step (1), the columns in which $A_{2,2}$ is located become singletons. We use these columns to perform singleton-deletions.

$\dots$

(i) Since $A_{1,i}, A_{2,i},\dots, A_{i,i}$ are all-zero matrices, after steps (1)-(i-1), the columns in which $A_{i,i}$ is located become singletons. We use these columns to perform singleton-deletions.

$\dots$

(s) Since $A_{1,s}, A_{2,s},\dots, A_{s,s}$ are all-zero matrices, after steps (1)-(s-1), the columns in which $A_{s,s}$ is located become singletons. We use these columns to perform singleton-deletions.

\end{framed}

We have eliminated all the $\Delta$'s in the last column, so the last column of $A$ becomes a singleton.
We use the last column to perform singleton-deletion, then the last row of $A$ is eliminated, and the second last column becomes the last \textbf{nonzero} column.

After the aforementioned singleton-deletions, the column $A^{\{z_1\}}$ must be a singleton.
Considering the submatrices where $\blacktriangle$'s are located as shown in the matrix $A$, we can deduce that in the last column of each such submatrix, $\blacktriangle$ is the only possible entry to take nonzero value, based on the facts that $A_{i,j}$ are all-zero matrices for $1\leq i < j\leq s$ and $A_{\{i\}}^{\{j\}}=A_{\{i+1\}}^{\{j+1\}}$ for $1\leq i, j\leq m-1$. So from Remark~\ref{rem:deletion}, the only possible obstacle to eliminating the remaining $\Delta$'s in the second last column of the matrix $A$ are the entries highlighted by $\blacktriangle$'s. Similar to the procedure described in the above frame, the remaining $\Delta$'s in the second last column of $A$ can be eliminated using a singleton procedure started by $A^{\{z_1\}}$. Then the second last column of $A$ becomes a singleton, and its corresponding row, i.e., the second last row of $A$, can be eliminated by related singleton-deletion.

In fact, we can repeat this procedure of making the last nonzero column become singleton by eliminating its $\Delta$'s at least $z_1$ times. Since $t_0\leq z_1$, we can eliminate all the last $t_0$ rows of $A$ by singleton-deletions. Once the last $t_0$ rows of $A$ were eliminated, the resulting submatrix is an upper-triangle matrix with a trivial successful singleton procedure, together with the singleton-deletions eliminating the last $t_0$ rows of $A$, we get a successful singleton procedure for all the rows of $A$.

\textbf{Claim 2}: For each $1\leq i\leq s$, if $\mathcal{S}_{i} \subseteq_{0} ((A^{\{m\}})^{T})^{-1}$, then $A_{i,i},A_{i,i+1},\dots,A_{i,s}$ are all-zero matrices.

From property (a), i.e., $A_{\{i\}}^{\{j\}}=A_{\{i+1\}}^{\{j+1\}}$ for $1\leq i, j\leq m-1$, we know that the part beyond the main diagonal of $A$ can be determined by the first row of $A$. For simplicity, we denote by $\mathbf{r}$ the first row of $A$.
For each $ i\in [s]$, if $\mathcal{S}_i \subseteq_{0} \mathbf{r}$, by property (a), we can check that $A_{i,i},A_{i,i+1},\dots,A_{i,s}$ are all-zero matrices. Since $\mathbf{r}= ((A^{\{m\}})^{T})^{-1}$, the result follows.

The proof of Lemma~\ref{lem:main} is completed.

\subsection{Main Theorem}
In this subsection, we provide a sufficient condition for bounding the minimum distance of cyclic codes with defining sets of a special type based on Lemma~\ref{lem:main}, which serves as an origin of the derivation of new bounds.

\begin{theorem}\label{thm:generalbound}
Let $A$ be an $m\times m$ matrix over $\mathcal{U}$ that satisfies the conditions of Lemma~\ref{lem:main} with $t_0=z_1$. Let $\mathcal{C}$ be a cyclic code with defining set $S_{\mathcal{C}}$. If $R(n, S_{\mathcal{C}})$ contains a consecutive part of
\begin{equation}\label{eq:s0}
    (0)^{y_s+z_{s+1}-z_1}(\Delta)^{1}(0)^{z_{s+1}}(\Delta)^{t_s}(0)^{z_s}(\Delta)^{t_{s-1}}(0)^{z_{s-1}}\dots(\Delta)^{t_1}(0)^{z_1},
\end{equation}
then
$$d(\mathcal{C})\geq y_s+z_{s+1}+1.$$
\end{theorem}

\begin{IEEEproof}
From Theorem~\ref{thm:schuab}, it is sufficient to show that:
$$\min\{{\rm prk}((M(\mathbf{u}))~|~\mathbf{u} \in \mathcal{A}(R(n, S_{\mathcal{C}})) \}\geq y_s+z_{s+1}+1.$$
So we need to show that for any $\mathbf{u} \in \mathcal{A}(R(n, S_{\mathcal{C}}))$, prk$(M(\mathbf{u}))\geq y_s+z_{s+1}+1$.

From Lemma~\ref{lem:shift}, without loss of generality, we assume that $\mathbf{u}$ has the following form:
$$(\overbrace{0,\dots,0}^{y_s+z_{s+1}-z_1},\Delta,\overbrace{ 0,\dots,0}^{z_{s+1}}, \overbrace{\Delta,\dots,\Delta}^{t_s},\overbrace{ 0,\dots,0}^{z_s}, \dots \overbrace{\Delta,\dots,\Delta}^{t_1},\overbrace{0,\dots,0}^{z_1},\dots)$$
in which we put $\Delta$ in positions where we do not know if there is a $0$
or a $\Delta^{+}$.

For a given $\mathbf{u} \in \mathcal{A}(R(n, S_{\mathcal{C}}))$, we have two cases as follows:
\begin{itemize}
  \item[(i)] $\mathbf{u}[y_s+z_{s+1}-z_1+1] = 0$,
  \item[(ii)] $\mathbf{u}[y_s+z_{s+1}-z_1+1] = \Delta^{+}$.
\end{itemize}

In case (i), $\mathbf{u}$ contains a consecutive zero sequence with length at least $y_s+2z_{s+1}-z_1+1$, from Lemma~\ref{lem:bch}, ${\rm prk}((M(\mathbf{u}))\geq y_s+2z_{s+1}-z_1+2>y_s+z_{s+1}+1$.
So, we only consider the case (ii).

When $\mathbf{u}[y_s+z_{s+1}-z_1+1] = \Delta^{+}$, it is easy to check that the first $y_s+z_{s+1}+1$ rows and columns with indices in $[y_s+z_{s+1}-z_1+1,2y_s+2z_{s+1}-z_1+1]$ of $M(\mathbf{u})$ is exactly the square matrix $A$, which has pseudo-rank $y_s+z_{s+1}+1$, so ${\rm prk}(M(\mathbf{u}))\geq y_s+z_{s+1}+1$.
The proof is completed.
\end{IEEEproof}

\begin{remark}
In fact, when $t_0<z_1$, if $R(n, S_{\mathcal{C}})$ contains a consecutive part of
\begin{equation}\label{eq:s1}
    (0)^{y_s+z_{s+1}-t_0}(\Delta)^{1}(0)^{z_{s+1}}(\Delta)^{t_s}(0)^{z_s}(\Delta)^{t_{s-1}}(0)^{z_{s-1}}\dots(\Delta)^{t_1}(0)^{z_1},
\end{equation}
from Lemma~\ref{lem:main} we can also deduce that $d(\mathcal{C})\geq y_s+z_{s+1}+1$.
But notice that if $R(n, S_{\mathcal{C}})$ contains a consecutive part of \eqref{eq:s1}, it also contains a consecutive part of \eqref{eq:s0}, hence we only need to consider the case of $t_0=z_1$, which is the circumstance Theorem~\ref{thm:generalbound} describes.

\end{remark}

According to Theorem~\ref{thm:generalbound}, we can obtain new bounds for the minimum distance of cyclic codes by carefully designing the defining sets. In the following subsection, we will address the problem of how to create meaningful defining sets to derive new bounds.

\subsection{General Bounds Derived from Theorem~\ref{thm:generalbound}}

In this subsection, we consider a specific case of Theorem~\ref{thm:generalbound}, which is $z_{s+1}+1=x_s=x_{s-1}=\dots=x_1=\delta$. Then we have

\[
\begin{aligned}
((A^{\{m\}})^{T})^{-1}&=((\Delta^{+})^{1}(0)^{\delta-1} (\Delta)^{t_s}(0)^{\delta-t_s}\cdots(\Delta)^{t_2}(0)^{\delta-t_{2}}(\Delta)^{t_1}(0)^{\delta-t_{1}}),\\
\mathcal{S}_{1}&=((\Delta^{+})^{1}(0)^{\delta-1}(\Delta)^{t_1-t_{2}+1}(0)^{\delta-(t_1-t_{2}+1)}
\cdots(\Delta)^{t_1-t_{s}+1}(0)^{\delta-(t_1-t_{s}+1)}(\Delta)^{\delta}),\\ \mathcal{S}_{2}&=((\Delta^{+})^{1}(0)^{\delta-1}(\Delta)^{t_2-t_{3}+1}(0)^{\delta-(t_2-t_{3}+1)}
\cdots(\Delta)^{t_2-t_{s}+1}(0)^{\delta-(t_2-t_{s}+1)}(\Delta)^{2\delta}), \\
& ~~ \vdots \\
\mathcal{S}_{s-1}&=((\Delta^{+})^{1}(0)^{\delta-1}(\Delta)^{t_{s-1}-t_{s}+1}(0)^{\delta-(t_{s-1}-t_{s}+1)}
(\Delta)^{(s-1)\delta}),\\
\mathcal{S}_{s}&=((\Delta^{+})^{1}(0)^{\delta-1}(\Delta)^{s\delta}).
\end{aligned}
\]
As we can see $\mathcal{S}_s\subseteq_{0} ((A^{\{m\}})^{T})^{-1}$ is trivial.
From $\mathcal{S}_1\subseteq_{0} ((A^{\{m\}})^{T})^{-1}$, we have the following $s-1$ inequalities

$$
\left\{
\begin{aligned}
t_s&\leq t_1-t_2+1,\\
t_{s-1}&\leq t_{1}-t_3+1,\\
& ~~ \vdots \\
t_3&\leq t_{1}-t_{s-1}+1,\\
t_2 &\leq t_{1}-t_{s}+1.
\end{aligned}
\right.
$$
Removing the duplicate inequalities, we obtain the following $\lceil\frac{s-1}{2}\rceil$ inequalities
$$
\left\{
\begin{aligned}
t_2+t_s-1&\leq t_1,\\
t_3+t_{s-1}-1&\leq t_{1},\\
& ~~ \vdots \\
t_{\lceil\frac{s+1}{2}\rceil}+t_{\lfloor\frac{s+3}{2}\rfloor}-1 &\leq t_{1}.
\end{aligned}
\right.
$$

Similarly, we can verify that for each $2\leq i\leq s-1$, from $\mathcal{S}_i\subseteq_{0} ((A^{\{m\}})^{T})^{-1}$, we can get a system of $\lceil\frac{s-i}{2}\rceil$ inequalities as follows

\begin{equation}\label{eq:main}
\left\{
\begin{aligned}
t_{i+1}+t_s-1 &\leq t_i,\\
t_{i+2}+t_{s-1}-1&\leq t_{i},\\
& ~~ \vdots \\
t_{\lceil\frac{s+i}{2}\rceil}+t_{\lfloor\frac{s+2+i}{2}\rfloor}-1 &\leq t_{i}.
\end{aligned}
\right.
\end{equation}

So, designing new meaningful patterns of defining sets is equivalent to finding solutions $(t_1,t_2,\dots,t_{s})$ of the system composed of inequalities in \eqref{eq:main} for all $1\leq i\leq s-1$, along with $t_0= z_1=\delta-t_1$.

We can run through all the inequalities in the order of $t_s,t_{s-1},\dots,t_1,t_0$ as follows
\begin{itemize}
  \item $\delta-1\geq t_s\geq 1$,
  \item $\delta-1\geq t_ {s-1}\geq 2t_s -1$,
  \item $\delta-1\geq t_{s-2}\geq t_{s}+t_{s-1}-1$,
  \item $\delta-1\geq t_{s-3}\geq \max\{t_{s}+t_{s-2}-1, 2t_{s-1}-1\}$,
  \item $\delta-1\geq t_{s-4}\geq \max\{t_{s}+t_{s-3}-1, t_{s-1}+t_{s-2}-1\}$,
  \item[] ~~~~~~~~$\vdots$
  \item $\delta-1\geq t_2\geq \max\{t_{s}+t_3-1,t_{s-1}+t_4-1,\dots, t_{\lceil\frac{s+2}{2}\rceil}+t_{\lfloor\frac{s+4}{2}\rfloor}-1 \}$,
  \item $\delta-1 \geq t_1\geq \max\{t_{s}+t_{2}-1,t_{s-1}+t_{3}-1,\dots,t_{\lceil\frac{s+1}{2}\rceil}+t_{\lfloor\frac{s+3}{2}\rfloor}-1\}$,
 \item $\delta-t_1 =t_0\geq 1$.
\end{itemize}

Formally, we have the following corollary from Theorem~\ref{thm:generalbound}.
\begin{corollary}\label{cor:bound0}
Let $\mathcal{C}$ be a cyclic code with defining
set $S_{\mathcal{C}}$.
Suppose that there are $\delta \in \mathbb{Z}$ with $\delta\geq 2$, $s\in \mathbb{Z}$ with $s\geq 1$ and $t_i\in \mathbb{Z}$ with $1\leq t_i\leq \delta-1$ for all $i\in [s]$  such that $R(n, S_{\mathcal{C}})$ contains a consecutive part of
$$(0)^{s\delta+t_1-1}(\Delta)^{1}(0)^{\delta-1}(\Delta)^{t_s}(0)^{\delta-t_s}(\Delta)^{t_{s-1}}(0)^{\delta-t_{s-1}}\dots(\Delta)^{t_1}(0)^{\delta-t_1}.$$

If $t_s,t_{s-1},\dots,t_1$ satisfy the following system of inequalities
\begin{equation}\label{eq:OO}
\left\{ \begin{aligned}
 t_s &\geq 1,\\
 t_ {s-1} &\geq 2t_s -1,\\
 t_{s-2} &\geq t_{s}+t_{s-1}-1,\\
 t_{s-3} &\geq \max\{t_{s}+t_{s-2}-1, 2t_{s-1}-1\},\\
 t_{s-4} &\geq \max\{t_{s}+t_{s-3}-1, t_{s-1}+t_{s-2}-1\}, \\
 &~~\vdots  \\
 t_2&\geq \max\{t_{s}+t_3-1,t_{s-1}+t_4-1,\dots,t_{\lceil\frac{s+2}{2}\rceil}+t_{\lfloor\frac{s+4}{2}\rfloor}-1 \},\\
 t_1&\geq \max\{t_{s}+t_{2}-1,t_{s-1}+t_{3}-1,\dots,t_{\lceil\frac{s+1}{2}\rceil}+t_{\lfloor\frac{s+3}{2}\rfloor}-1\},
\end{aligned}
\right.
\end{equation}

then
$$d(\mathcal{C})\geq (s+1)\delta.$$

\end{corollary}

\begin{remark}\label{rem:solution}
From Corollary~\ref{cor:bound0}, once we find a solution $(t_s,t_{s-1},\dots,t_1)$ to the system of inequalities \eqref{eq:OO}, we can obtain a specific bound for the minimum distance of cyclic codes.
In general, the number of solutions may be huge when $\delta$ is large and $s$ is small.
Here we provide three types of solutions with general patterns.

\textbf{Type I:}
\begin{equation}\label{eq:solution01}
(t_s,t_{s-1},\dots,t_1)=(1,1,\dots,1,x), \mbox{ where } 1\leq x\leq \delta-1 .
\end{equation}

\textbf{Type II:}
\begin{equation}\label{eq:solution02}
(t_s,t_{s-1},\dots,t_1)=
\left\{ \begin{array}{l}
(1,2,3,\dots,s), \mbox{ when } s \leq \delta-1, \\
(2,3,4,\dots,s+1), \mbox{ when }  s+1 \leq \delta-1, \\
(3,4,5,\dots,s+2), \mbox{ when }  s+2 \leq \delta-1, \\
 ~~\vdots  \\
\end{array}
\right.
\end{equation}

\textbf{Type III:}
\begin{equation}\label{eq:solution03}
(t_s,t_{s-1},\dots,t_1)=
\left\{ \begin{array}{l}
(1,2,2,3,3,4,4,\dots,\lceil\frac{s-1}{2}\rceil+1)), \mbox{ when } \lceil\frac{s-1}{2}\rceil+1 \leq \delta-1, \\
(1,3,3,5,5,7,7,\dots,2\lceil\frac{s-1}{2}\rceil+1), \mbox{ when } 2\lceil\frac{s-1}{2}\rceil+1\leq \delta-1, \\
(1,4,4,7,7,10,10,\dots,3\lceil\frac{s-1}{2}\rceil+1), \mbox{ when } 3\lceil\frac{s-1}{2}\rceil+1\leq \delta-1, \\
~~ \vdots  \\
\end{array}
\right.
\end{equation}

There are some other solutions that may not have general forms. For example, let $s=7$, $\delta=10$, the solutions to the system of inequalities \eqref{eq:OO} could be
$$(t_s,t_{s-1},\dots,t_1)=
\left\{ \begin{array}{l}
(1,1,1,2,2,3,5), \\
(1,2,3,4,8,9,9), \\
(2,2,3,4,5,8,9),  \\

~~ \vdots  \\
\end{array}
\right.$$

\end{remark}

From the three types of general solutions \eqref{eq:solution01}, \eqref{eq:solution02} and \eqref{eq:solution03} in Remark~\ref{rem:solution}, we obtain the following three explicit bounds.

\begin{theorem}\label{thm:explicitbound01} (Bound I)
Let $\mathcal{C}$ be a cyclic code with defining set $S_{\mathcal{C}}$.
Suppose that there are integers $\delta, s, x$ with $\delta\geq 2$, $s\geq 1$ and $1\leq x\leq \delta-1$ such that $R(n, S_{\mathcal{C}})$ contains a consecutive part of
 $$(0)^{s\delta+x-1}(\Delta)^{1}((0)^{\delta-1}(\Delta)^{1})^{s}(\Delta)^{x}(0)^{\delta-x},$$
then
$$d\geq (s+1)\delta.$$

\end{theorem}

\begin{remark}
Bound I degenerates to the Betti-Sala bound when we take $x=1$ in Theorem~\ref{thm:explicitbound01}. So Theorem~\ref{thm:explicitbound01} is a generalization of the Betti-Sala bound.
\end{remark}

\begin{theorem}\label{thm:explicitbound02} (Bound II)
Let $\mathcal{C}$ be a cyclic code with defining
set $S_{\mathcal{C}}$.
Suppose that there are integers $\delta, s, y$ with $\delta\geq 2$, $s\geq 1$ and $1\leq y\leq \delta-s$ such that $R(n, S_{\mathcal{C}})$ contains a consecutive part of
 $$(0)^{s(\delta+1)+y-2}(\Delta)^{1}(0)^{\delta-1}(\Delta)^{y}(0)^{\delta-y}(\Delta)^{y+1}(0)^{\delta-y-1}\dots(\Delta)^{y+s-1}(0)^{\delta-y-s+1},$$
then
$$d\geq (s+1)\delta.$$

\end{theorem}

\begin{theorem}\label{thm:explicitbound03} (Bound III)
Let $\mathcal{C}$ be a cyclic code with defining
set $S_{\mathcal{C}}$.
Suppose that there are integers $\delta, s, t$ with $\delta\geq 2$, $s\geq 1$ and $0\leq t\lceil\frac{s-1}{2}\rceil\leq \delta-2$ such that $R(n, S_{\mathcal{C}})$ contains a consecutive part of
  $$(0)^{s\delta+t\lceil\frac{s-1}{2}\rceil} ((\Delta)^{1}(0)^{\delta-1})^{2} ((\Delta)^{t+1}(0)^{\delta-t-1})^{2} ((\Delta)^{2t+1}(0)^{\delta-2t-1})^{2}\dots(\Delta)^{\lceil\frac{s-1}{2}\rceil t+1}(0)^{\delta-\lceil\frac{s-1}{2}\rceil t-1},$$
then
$$d\geq (s+1)\delta.$$

\end{theorem}

We provide illustrative examples to demonstrate that Bounds I, II and III derived from Corollary~\ref{cor:bound0}, can occasionally serve as tight bounds, outperforming both the Betti-Sala bound and the BCH bound.

\begin{example}

\begin{itemize}

    \item[(1)] Let $\mathcal{C}$ be a $25$-ary $[n,k,d]$ cyclic code with $n=24$, $k=7$, and defining set
$$S_{\mathcal{C}}=\{1,2,3,4,5,6,7,8,9,10,12,13,15,16,18,19,22\}.$$

From the BCH bound, $d\geq 11$. From the Betti-Sala bound, $d\geq 9$. Following the notation of Bound I, we can check that $s=3$, $\delta=3$, $x=2$, thus we obtain that $d\geq 12$ by Bound I. In addition, by Magma software, the minimum distance $d$ is actually equal to $12$.

    \item[(2)] Let $\mathcal{C}$ be a $31$-ary $[n,k,d]$ cyclic code with $n=30$, $k=9$, and defining set
$$S_{\mathcal{C}}=\{1,2,\dots,12,14,15,16,17,20,21,22,26,27\}.$$

From the BCH bound, $d\geq 13$. The condition of the Betti-Sala bound is not satisfied. Following the notation of Bound II, we can check that $s=2$, $\delta=5$, $y=2$, thus we obtain that $d\geq 15$ by Bound II. In addition, by Magma software, the minimum distance $d$ is actually equal to $15$.

    \item[(3)] Let $\mathcal{C}$ be a $37$-ary $[n,k,d]$ cyclic code with $n=36$, $k=14$, and defining set
$$S_{\mathcal{C}}=\{1,2,\dots,14,16,17,18,20,21,22,26,30\}.$$

From the BCH bound, $d\geq 15$. From the Betti-Sala bound, $d\geq 12$. Following the notation of Bound III, we can check that $s=3$, $\delta=4$, $t=2$, thus we obtain that $d\geq 16$ by Bound III. In addition, by Magma software, the minimum distance $d$ is actually equal to $16$.

\end{itemize}

\end{example}

As noted in Remark~\ref{rem:solution}, there may be other solutions to the system of inequalities \eqref{eq:OO} that do not have the general forms as Types I, II, and III.
We provide an illustrative example to demonstrate that the bounds derived from such solutions can also occasionally be tight, outperforming both the Betti-Sala bound and the BCH bound.

\begin{example}
 Let $\mathcal{C}$ be a $31$-ary $[n,k,d]$ cyclic code with $n=30$, $k=9$, and defining set
$$S_{\mathcal{C}}=\{1,2,3,4,5,6,7,8,9,10,11,12,13,15,16,17,18,21,22,23,28\}.$$
From the BCH bound, $d\geq 14$. The condition of the Betti-Sala bound is not satisfied. Following the notation of Corollary~\ref{cor:bound0}, we can check that $s=2$, $\delta=5$, $t_2=2$, $t_1=4$, thus we obtain that $d\geq 15$ by Corollary~\ref{cor:bound0}. In addition, by Magma software, the minimum distance $d$ is actually equal to $15$.
\end{example}

\section{A Generalization for a Special Case in the Main Theorem}
In this section, we present a generalization of Theorem \ref{thm:generalbound} for the case of $s=1$. As a result, we derive a corollary that will play a crucial role in the constructions of distance-optimal LRCs in the subsequent section.

Firstly, for the convenience of the reader, we restate the result of Theorem \ref{thm:generalbound} when $s=1$.
\begin{theorem}\label{thm:2delta-02}
Let $\mathcal{C}$ be a cyclic code with defining set $S_{\mathcal{C}}$.
Suppose that there are $\delta \in \mathbb{Z}$ with $\delta\geq 2$, $t\in \mathbb{Z}$ with $1\leq t\leq \delta-1$ such that $R(n, S_{\mathcal{C}})$ contains a consecutive part of
$$(0)^{\delta+t-1}(\Delta)^{1}(0)^{\delta-1}(\Delta)^{t}(0)^{\delta-t}.$$
Then
$$d(\mathcal{C})\geq 2\delta.$$

\end{theorem}

Next, we give a generalization of Theorem \ref{thm:2delta-02} as follows.

\begin{theorem}\label{thm:2delta-02-long}
Let $\mathcal{C}$ be a cyclic code with defining set $S_{\mathcal{C}}$.
Suppose that there are integers $\delta\geq 2$, $1\leq t \leq \delta-1$, $m\geq 0$ such that $R(n, S_{\mathcal{C}})$ contains a consecutive part of
$$(0)^{\delta+t-1}(\Delta)^{1}((0)^{\delta-1}(\Delta)^{2})^{m}(0)^{\delta-1}(\Delta)^{t}(0)^{\delta-t},$$
then
$$d(\mathcal{C})\geq 2\delta.$$

\end{theorem}

\begin{IEEEproof}
When $m=0$, this theorem degenerates to Theorem \ref{thm:2delta-02}. In the rest of the proof, we suppose $m\geq 1$.
From Theorem~\ref{thm:schuab}, it is sufficient to show that:
$$\min\{{\rm prk}((M(\mathbf{u}))~|~\mathbf{u} \in \mathcal{A}(R(n, S_{\mathcal{C}})) \}\geq 2\delta,$$
which is equivalent to prove that for any $\mathbf{u} \in \mathcal{A}(R(n, S_{\mathcal{C}}))$, ${\rm prk}(M(\mathbf{u}))\geq 2\delta$.

From Lemma~\ref{lem:shift}, without loss of generality, we assume that $\mathbf{u}$ has the following form:
$$(\overbrace{0,\dots,0}^{\delta+t-1},\Delta,\overbrace{
\underline{\overbrace{0,\dots,0}^{\delta-1},\Delta,\Delta},\dots, \underline{\overbrace{0,\dots,0}^{\delta-1},\Delta,\Delta}}^{\mbox{$m$ blocks}}, \overbrace{0,\dots,0}^{\delta-1},\overbrace{\Delta,\dots,\Delta}^{t},\overbrace{0,\dots,0}^{\delta-t},\dots)$$
in which we put $\Delta$ in positions where we do not know if there is a $0$
or a $\Delta^{+}$.

For a given $\mathbf{u} \in \mathcal{A}(R(n, S_{\mathcal{C}}))$, we have two cases as follows:
\begin{itemize}
  \item[(i)] $\mathbf{u}[\delta+t] = 0$,
  \item[(ii)] $\mathbf{u}[\delta+t] = \Delta^{+}$.
\end{itemize}

In case (i), $\mathbf{u}$ contains a consecutive zero sequence with length at least $2\delta+t-1$, from Lemma~\ref{lem:bch}, ${\rm prk}((M(\mathbf{u}))\geq 2\delta+t$. So, we only consider the case (ii) in the following.

There are $m$ pairs of $(\Delta,\Delta)$ among the $m$ blocks in $\mathbf{u}$. For $1\leq i\leq m$, we denote by $(\Delta,\Delta)_i$ the $i$-th pair.
Recall that $\Delta$ could be $0$ or $\Delta^{+}$, here we only care about the second component of $(\Delta,\Delta)_i$. Let $j$ be the \textbf{first} index such that $(\Delta,\Delta)_j=(\Delta,\Delta^{+})$, then we have the following three subcases:
\begin{itemize}
  \item[(1)] $j=0$;
  \item[(2)] $1\leq j\leq m-1$;
    \item[(3)] $j=m$.
\end{itemize}

We need to prove that for all these three subcases, ${\rm prk}((M(\mathbf{u}))\geq 2\delta$, i.e., there is a successful singleton procedure for a $2\delta\times 2\delta$ submatrix of $M(\mathbf{u})$.

\textbf{Subcase} (1):
In this subcase, $(\Delta,\Delta)_i=(\Delta,0)$ for all $i\in [m]$, then $\mathbf{u}$ has the form
$$(\overbrace{0,\dots,0}^{\delta+t-1},\Delta^{+},\overbrace{
\underline{\overbrace{0,\dots,0}^{\delta-1},\Delta,0},\dots, \underline{\overbrace{0,\dots,0}^{\delta-1},\Delta,0}}^{\mbox{$m$ blocks}}, \overbrace{0,\dots,0}^{\delta-1},\overbrace{\Delta,\dots,\Delta}^{t},\overbrace{0,\dots,0}^{\delta-t},\dots).$$

Let $\mathcal{I}_{0}=[\delta+t,3\delta+t-1]$, the first $2\delta$ rows of $M(\mathbf{u})^{\mathcal{I}_0}$ form the following $2\delta \times 2\delta$ matrix $X$,

\[\small
\begin{tabular}{r}
$\delta \left\{ \lefteqn{\phantom{\begin{matrix}  a_1 \\ a_2 \\ a_3 \\ a_4 \\ a_5 \\ a_6  \end{matrix}}} \right.$\\ 
$t \left\{ \lefteqn{\phantom{\begin{matrix}    b_0\\ b1 \\ b_3 \end{matrix}}} \right.$ \\
$\delta-t \left\{ \lefteqn{\phantom{\begin{matrix}    b_0\\ b1 \\ b_3 \end{matrix}}} \right.$
\end{tabular} 
\hspace{-1em} 
\left[\phantom{\begin{matrix} a_1 \\ a_2 \\ a_3 \\ a_4 \\ a_5 \\ a_6 \\ a_7 \\ a_8 \\ a_9 \\ a_{10} \\a_{11} \\a_{12} \end{matrix}}
\right.\hspace{-1.5em}
\begin{tabular}{CCCCCC|CCCCCC}
\Delta^{+}&&&&&&\Delta&&&&& \\
&\Delta^{+}&&&&&&\Delta&&&&  \\
&&\Delta^{+}&&&&&&\Delta&&&  \\
&&&\ddots&&&&&&\ddots&&    \\
&&&&\Delta^{+}&&&&&&\Delta& \\
&&&&&\Delta^{+}&&&&&&\Delta   \\ \hline
&&&&&&\Delta^{+}&&&&&    \\
&&&&&&&\Delta^{+}&&&&  \\
&&&&&&& &\Delta^{+}&&&   \\
\Delta&&&&&&&&&\ddots&    \\
\vdots&\ddots&&&&&&&&& \Delta^{+} &    \\
\Delta&\cdots&\Delta&&&&&&&&  & \Delta^{+}
\end{tabular} \hspace{-1.5em}
\left.\phantom{\begin{matrix} a_1 \\ a_2 \\ a_3 \\ a_4 \\ a_5 \\ a_6 \\ a_7 \\ a_8 \\ a_9 \\ a_{10} \\a_{11} \\a_{12} \end{matrix}}\right]
=X,\]
where all blank cells in the matrix are represented as ``0''.

It can be easily checked that matrix $X$ satisfies the conditions in Lemma~\ref{lem:main}, hence ${\rm prk}(X)= 2\delta$.

\textbf{Subcase} (2):
In this subcase, $\mathbf{u}$ has the form
$$(\overbrace{0,\dots,0}^{\delta+t-1},\mathbf{\Delta^{+}},
\overbrace{\underline{\overbrace{\mathbf{0},\dots,\mathbf{0}}^{\delta-1},\Delta,0},\dots, \underline{\overbrace{0,\dots,0}^{\delta-1},\Delta,0}}^{\mbox{$(j-1)$ blocks}},
\underline{\overbrace{0,\dots,0}^{\delta-1},\Delta,\Delta^{+}},
\underline{\overbrace{0,\dots,0}^{\delta-1},\mathbf{\Delta},\mathbf{0}},
\overbrace{\mathbf{0},\dots,\mathbf{0}}^{\delta-2},\dots).$$

We choose $\mathcal{I}_j=[\delta+t,2\delta+t-1]\cup [(j+1)(\delta+1)+\delta+t-1,(j+1)(\delta+1)+2\delta+t-2]$ (as bolded). We can check that for each $j\in [m-1]$, the first $2\delta$ rows of $M(\mathbf{u})^{\mathcal{I}_j}$ form the following matrix $Y$,

\[\small
\begin{tabular}{r}
$\delta \left\{ \lefteqn{\phantom{\begin{matrix}  a_1 \\ a_2 \\ a_3 \\ a_4 \\ a_5 \\ a_6  \end{matrix}}} \right.$\\ 
$t \left\{ \lefteqn{\phantom{\begin{matrix}    b_0\\ b1 \\ b_3 \end{matrix}}} \right.$ \\
$\delta-t \left\{ \lefteqn{\phantom{\begin{matrix}    b_0\\ b1 \\ b_3 \end{matrix}}} \right.$
\end{tabular} 
\hspace{-1em} 
\left[\phantom{\begin{matrix} a_1 \\ a_2 \\ a_3 \\ a_4 \\ a_5 \\ a_6 \\ a_7 \\ a_8 \\ a_9 \\ a_{10} \\a_{11} \\a_{12} \end{matrix}}
\right.\hspace{-1.5em}
\begin{tabular}{CCCCCC|CCCCCC}
\Delta^{+}&&&&&&\Delta&&&&& \\
&\Delta^{+}&&&&&&\Delta&&&&  \\
&&\Delta^{+}&&&&&&\Delta&&&  \\
&&&\ddots&&&&&&\ddots&&    \\
&&&&\Delta^{+}&&&&&&\Delta& \\
&&&&&\Delta^{+}&&&&&&\Delta   \\ \hline
&&&&&&\Delta^{+}&&&&&    \\
&&&&&&\Delta&\Delta^{+}&&&&  \\
&&&&&&&\Delta &\Delta^{+}&&&   \\
\Delta&&&&&&&&\ddots&\ddots&    \\
\vdots&\ddots&&&&&&&&\Delta& \Delta^{+} &    \\
\Delta&\cdots&\Delta&&&&&&&& \Delta & \Delta^{+}
\end{tabular} \hspace{-1.5em}
\left.\phantom{\begin{matrix} a_1 \\ a_2 \\ a_3 \\ a_4 \\ a_5 \\ a_6 \\ a_7 \\ a_8 \\ a_9 \\ a_{10} \\a_{11} \\a_{12} \end{matrix}}\right]
=Y,\]
where all blank cells in the matrix are represented as ``0''.

Applying the ideas used in proving Lemma~\ref{lem:main} (recursively eliminating the $\Delta$'s in the last nonzero column to make the last nonzero column become a singleton), one can check that there exists a successful singleton procedure for all the rows of $Y$ in the following order:
\begin{align*}
Y^{\{\delta\}}\rightarrow & Y^{\{2\delta\}} \rightarrow  \\
Y^{\{\delta-1\}} \rightarrow & Y^{\{2\delta-1\}} \rightarrow\\
Y^{\{\delta-2\}} \rightarrow & Y^{\{2\delta-2\}} \rightarrow\\
\dots\\
Y^{\{2\}}\rightarrow & Y^{\{\delta+2\}} \rightarrow\\
Y^{\{1\}}\rightarrow & Y^{\{\delta+1\}}
\end{align*}

\textbf{Subcase} (3):
When $j=m$, $\mathbf{u}$ has the form
$$(\overbrace{0,\dots,0}^{\delta+t-1},\mathbf{\Delta^{+}},
\overbrace{\underline{\overbrace{\mathbf{0},\dots,\mathbf{0}}^{\delta-1},\Delta,0},\dots, \underline{\overbrace{0,\dots,0}^{\delta-1},\Delta,0}}^{\mbox{$(m-1)$ blocks}},
\underline{\overbrace{0,\dots,0}^{\delta-1},\Delta,\Delta^{+}},
\overbrace{0,\dots,0}^{\delta-1},
\overbrace{\mathbf{\Delta},\dots,\mathbf{\Delta}}^{t},
\overbrace{\mathbf{0},\dots,\mathbf{0}}^{\delta-t},\dots).$$

We choose $\mathcal{I}_m=[\delta+t,2\delta+t-1]\cup [(m+1)(\delta+1)+\delta+t-1,(m+1)(\delta+1)+2\delta+t-2]$ (as bolded), the first $2\delta$ rows of $M(\mathbf{u})^{\mathcal{I}_m}$ form the following $2\delta \times 2\delta$ matrix $Z$,

\[\small
\begin{tabular}{r}
$\delta \left\{ \lefteqn{\phantom{\begin{matrix}  a_1 \\ a_2 \\ a_3 \\ a_4 \\ a_5 \\ a_6  \end{matrix}}} \right.$\\ 
$t \left\{ \lefteqn{\phantom{\begin{matrix}    b_0\\ b1 \\ b_3 \end{matrix}}} \right.$ \\
$\delta-t \left\{ \lefteqn{\phantom{\begin{matrix}    b_0\\ b1 \\ b_3 \end{matrix}}} \right.$
\end{tabular} 
\hspace{-1em} 
\left[\phantom{\begin{matrix} a_1 \\ a_2 \\ a_3 \\ a_4 \\ a_5 \\ a_6 \\ a_7 \\ a_8 \\ a_9 \\ a_{10} \\a_{11} \\a_{12} \end{matrix}}
\right.\hspace{-1.5em}
\begin{tabular}{CCCCCC|CCCCCC}
\Delta^{+}&&&&&&\Delta&\cdots&\Delta&&& \\
&\Delta^{+}&&&&&&\Delta&&\Delta&&  \\
&&\Delta^{+}&&&&&&\ddots&&\ddots&  \\
&&&\ddots&&&&&&\Delta&&  \Delta  \\
&&&&\Delta^{+}&&&&&&\Delta&\vdots \\
&&&&&\Delta^{+}&&&&&&\Delta   \\ \hline
&&&&&&\Delta^{+}&&&&&    \\
&&&&&&\Delta&\Delta^{+}&&&&  \\
&&&&&&&\Delta &\Delta^{+}&&&   \\
\Delta&&&&&&&&\ddots&\ddots&    \\
\vdots&\ddots&&&&&&&&\Delta& \Delta^{+} &    \\
\Delta&\cdots&\Delta&&&&&&&& \Delta & \Delta^{+}
\end{tabular} \hspace{-1.5em}
\left.\phantom{\begin{matrix} a_1 \\ a_2 \\ a_3 \\ a_4 \\ a_5 \\ a_6 \\ a_7 \\ a_8 \\ a_9 \\ a_{10} \\a_{11} \\a_{12} \end{matrix}}\right]
\hspace{-1em}
\begin{tabular}{r}
$\lefteqn{ \phantom{\begin{matrix}    b_0\\ b1 \\ b_3 \end{matrix}}} $ \\
$\left. \lefteqn{\phantom{\begin{matrix}    b_0\\ b1 \\ b_3 \end{matrix}}} \right\}t$ \\
$ \lefteqn{\phantom{\begin{matrix}  a_1 \\ a_2 \\ a_3 \\ a_4 \\ a_5 \\ a_6  \end{matrix}}} $ 
\end{tabular}
\hspace{-1em}
=Z,\]
where all blank cells in the matrix are represented as ``0''.

Note that when $t=1$, matrix $Z$ coincides with matrix $Y$, so we assume $t\geq 2$ in the rest of the proof.
Similarly, there exists a successful singleton procedure for all the rows of $Z$ in the following order:
\begin{align}
\label{ord:21} (Z^{\{\delta-t+1\}},\dots, Z^{\{\delta\}}) \rightarrow &Z^{\{2\delta\}} \rightarrow \\
Z^{\{\delta-t\}} \rightarrow & Z^{\{2\delta-1\}} \rightarrow\\
Z^{\{\delta-t-1\}} \rightarrow & Z^{\{2\delta-2\}} \rightarrow\\
\dots\\
\label{ord:22} Z^{\{1\}}\rightarrow & Z^{\{\delta+t\}} \rightarrow\\
\label{ord:23} Z^{\{\delta+t-1\}}\rightarrow & Z^{\{\delta+t-2\}}\rightarrow\dots\rightarrow Z^{\{\delta+1\}}
\end{align}
where in \eqref{ord:21}, $(Z^{\{\delta-t+1\}},\dots, Z^{\{\delta\}})$ means that we can perform singleton-deletions simultaneously, as $Z^{\{\delta-t+1\}},\dots, Z^{\{\delta\}}$ originally are singletons.
After performing singleton-deletions in \eqref{ord:21}-\eqref{ord:22}, the resulting submatrix is a lower triangular matrix, which has a trivial successful singleton procedure as shown in \eqref{ord:23}.

In summary, for any $\mathbf{u} \in \mathcal{A}(R(n, S_{\mathcal{C}}))$, we can always find a $2\delta\times 2\delta$ submatrix of $M(\mathbf{u})$ which possess a successful singleton procedure for all its rows, i.e., ${\rm prk}(M(\mathbf{u}))= 2\delta$.
The proof is completed.
\end{IEEEproof}

\begin{example}

Suppose that $\delta=4$, $m=2$, $t=2$ in Theorem~\ref{thm:2delta-02-long}, from the proof of Theorem~\ref{thm:2delta-02-long}, we only need to prove that the following two $8\times 8$ matrices $Y$ and $Z$ have successful singleton procedures for all their rows.
\[
Y=\left[
  \begin{array}{cccc|cccc}
\Delta^{+}     & 0 &  0 & 0 & \Delta  &   0 & 0 & 0   \\
0 & \Delta^{+} & 0 &  0 & 0 & \Delta  & 0 & 0    \\
0  & 0  & \Delta^{+} & 0 &  0 &   0 & \Delta  & 0   \\
0  & 0  & 0 & \Delta^{+} & 0 &  0 &   0 & \Delta   \\ \hline
0  & 0  & 0 & 0 & \Delta^{+} & 0 &  0 &   0  \\
0  & 0  & 0 & 0 &  \Delta   & \Delta^{+} & 0 &  0  \\
\Delta  & 0 & 0  & 0 & 0 &  \Delta   & \Delta^{+} & 0    \\
\Delta  & \Delta & 0 & 0   & 0  & 0  &   \Delta & \Delta^{+}
  \end{array} \right]
    \begin{array}{c}
   {\rm 7th~ step} \\
   {\rm 5th~ step}\\
    {\rm 3rd~ step}\\
    {\rm 1st~ step}\\
    {\rm 8th~ step}\\
    {\rm 6th~ step} \\
    {\rm 4th~ step}\\
    {\rm 2nd~ step}
    \end{array}
\]

\[
Z=\left[
  \begin{array}{cccc|cccc}
\Delta^{+}     & 0 &  0 & 0 & \Delta  &   \Delta & 0 & 0   \\
0 & \Delta^{+} & 0 &  0 & 0 & \Delta  & \Delta & 0    \\
0  & 0  & \Delta^{+} & 0 &  0 &   0 & \Delta  & \Delta   \\
0  & 0  & 0 & \Delta^{+} & 0 &  0 &   0 & \Delta   \\ \hline
0  & 0  & 0 & 0 & \Delta^{+} & 0 &  0 &   0  \\
0  & 0  & 0 & 0 &  \Delta   & \Delta^{+} & 0 &  0  \\
\Delta  & 0 & 0  & 0 & 0 &  \Delta   & \Delta^{+} & 0    \\
\Delta  & \Delta & 0 & 0   & 0  & 0  &   \Delta & \Delta^{+}
  \end{array} \right]
    \begin{array}{c}
   {\rm 6th~ step} \\
   {\rm 4th~ step}\\
    {\rm 2rd~ step}\\
    {\rm 1st~ step}\\
    {\rm 8th~ step}\\
    {\rm 7th~ step} \\
    {\rm 5th~ step}\\
    {\rm 3nd~ step}
    \end{array}
\]

The singleton procedures can be successfully done by the order as indicated beside the matrices.
\end{example}

The following corollary deals with a specific case of $(\delta+1)\mid n$ and $t=1$ in Theorem~\ref{thm:2delta-02-long}. This particular case is crucial for our constructions in the next section.

\begin{corollary}\label{cor:boubdforlrc}
 Let $q$ be a prime power and $i_0$ be an integer. Suppose that $n\geq 2$, $\delta\geq 2$ are positive integers such that $\gcd(q,n)=1$ and $(\delta+1)\mid n$, define $\rho=\frac{n}{\delta+1}$. Let $\mathcal{C}$ be a cyclic code of length $n$ with defining set $L\cup D$, where
$L=\{i_0+i+j(\delta+1)~|~1\leq i\leq \delta-1, 0\leq j\leq \rho-1\}$,
$D=\{i_1,i_2\}$ for some $0\leq i_1< i_2\leq n-1$.
If $L \cap D=\emptyset$ and $i_1\not\equiv i_2(\bmod~ \delta+1)$, then $d(\mathcal{C})\geq 2\delta$.
\end{corollary}

\begin{IEEEproof}
From $L \cap D=\emptyset$, we have $\langle i_1\rangle_{\delta+1}=i_0$ or $\langle i_1\rangle_{\delta+1}=i_0-1$. When $i_2=i_1+1$, we can check that $\{\langle i_1-\delta+1\rangle_n, \dots, \langle i_1\rangle_n,\langle i_1+1\rangle_n,\dots, \langle i_1+\delta\rangle_n\} \subseteq L\cup D$, by the BCH bound, $d(\mathcal{C})\geq 2\delta+1$. In the following, we only consider the situation of $i_2> i_1+1$.

Without loss of generality, we assume $i_0=0$, and it is only need to consider the following two cases:
\begin{itemize}
  \item[(1)] $i_1=\delta$;
  \item[(2)] $i_1=\delta+1$.
\end{itemize}

\textbf{Case} (1): When $i_1=\delta$, due to the assumption of the corollary, $i_2=(\ell+1)(\delta+1)$ for some $\ell \in [\rho-1]$. For convenience, we write $R(n, S_{\mathcal{C}})$ column by column with column size $\delta+1$, then $R(n, S_{\mathcal{C}})$ has the form

\[ \delta+1\left\{
  \begin{array}{cccccccccccccccc}
0 & \bovermat{$\ell-1$}{0 & \cdots &0} & 0& \bovermat{$\rho-1-\ell$}{0 &\dots &0}\\
\vdots & \vdots & \vdots &\vdots &\vdots&\vdots &\vdots&\vdots\\
0 & 0 & \cdots &0 &0 &0 &\dots & 0\\
0 & 0 & \cdots &0 &0 &0 &\dots & 0\\
\hline
0 & \Delta & \cdots &\Delta &\Delta &\Delta &\dots & \Delta \\
\Delta & \Delta & \cdots&\Delta &0 &\Delta &\dots & \Delta
  \end{array} \right.
.\]
From which we can see $R(n, S_{\mathcal{C}})$ contains a consecutive part of
$$(0)^{\delta}(\Delta)^{1}((0)^{\delta-1}(\Delta)^{2})^{\ell-1}(0)^{\delta-1}(\Delta)^{1}(0)^{\delta-1}.$$ By Theorem~\ref{thm:2delta-02-long}, $d(\mathcal{C})\geq 2\delta$.

\textbf{Case} (2): When $i_1=\delta+1$, due to the assumption of the corollary, $i_2=\delta+\ell(\delta+1)$ for some $\ell \in [\rho-1]$. Similarly, in a sequential column-wise reading manner, $R(n, S_{\mathcal{C}})$ has the form

\[ \delta+1\left\{
  \begin{array}{cccccccccccccccc}
0 & \bovermat{$\ell-1$}{0 & \cdots &0} & 0& \bovermat{$\rho-\ell-1$}{0 &\dots &0}\\
\vdots & \vdots & \vdots &\vdots &\vdots&\vdots &\vdots&\vdots\\
0 & 0 & \cdots &0 &0 &0 &\dots & 0\\
0 & 0 & \cdots &0 &0 &0 &\dots & 0\\
\hline
\Delta & \Delta & \cdots &\Delta &0 &\Delta &\dots & \Delta \\
0& \Delta & \cdots&\Delta &\Delta &\Delta &\dots & \Delta
  \end{array} \right.
.\]
From which we can see $R(n, S_{\mathcal{C}})$ contains a consecutive part of
$$(0)^{\delta}(\Delta)^{1}((0)^{\delta-1}(\Delta)^{2})^{\rho-\ell-1}(0)^{\delta-1}(\Delta)^{1}(0)^{\delta-1}.$$ By Theorem~\ref{thm:2delta-02-long}, $d(\mathcal{C})\geq 2\delta$.

The proof is completed.
\end{IEEEproof}

\section{Applications to the Constructions of LRCs}

 In this section, we construct distance-optimal $(2,\delta)$-LRCs with unbounded length and minimum distance $2\delta$ by utilizing the newly derived bound for the minimum distance of (consta)cyclic codes. It turns out that the minimum distance of our constructions exceeds $r+\delta+1$.

For an $(r,\delta)$-LRC $\mathcal{C}$ with code length $n$, let $S_i\subset [n]$ be the repair set for
the $i$-th code symbol, we say that $\mathcal{C}$ has disjoint local repair groups if $ (r +\delta-1) \mid n$ and each subcode $\mathcal{C}|_{S_i}$ is an $[r +\delta-1, r,\delta]$ MDS code.
We will employ Lemma \ref{lem:locality} to construct cyclic and constacyclic LRCs with $(r, \delta)$-locality, which always have disjoint local repair groups.

Throughout this section, we focus on $r=2$ hence assume that $(\delta+1) \mid n$ and denote $\rho=\frac{n}{\delta+1}$.

The following lemma is useful for our discussion.
\begin{lemma}\label{lem:icc}(See \cite[Corollary~1]{Chen22})
Let $\mathcal{C}$ be a Singleton-optimal $[n,k,d]$ $(r, \delta)$-LRC
with disjoint local repair groups. If $d = 2\delta + 1$ and $r = 2$, then
$n \leq q + 1$.
\end{lemma}

From Lemma~\ref{lem:icc}, we can get the following corollary.
\begin{corollary}\label{cor:icc}
Let $\mathcal{C}$ be an $[n,k,d]$ $(2, \delta)$-LRC
with disjoint local repair groups. If $2\delta+1=n-k- \left(\left\lceil \frac{k}{2}\right\rceil -1 \right)(\delta-1) +1$ and $n> q+ 1$, then
$d \leq 2\delta$.
\end{corollary}
\begin{IEEEproof}
  From the Singleton-like bound \eqref{eq:singleton}, we have $d\leq 2\delta+1$. If $d=2\delta+1$, i.e., $\mathcal{C}$ is Singleton-optimal, from Lemma~\ref{lem:icc}, $n\leq q+1$, contradiction. So $d \leq 2\delta$.
\end{IEEEproof}

Our goal is to construct LRCs with unbounded length, thus we assume $n>q+1$ in the following.

For the case of $(\delta+1)\mid (q-1)$, we have the following construction of distance-optimal LRCs with unbounded length from cyclic codes.
\begin{theorem}\label{thm:olrcq-1}
Let $q$ be a prime power and $\delta\geq 2$ be an integer such that $(\delta+1)\mid (q-1)$. If $\gcd(\rho,\delta+1)=1$, then there exists a distance-optimal $(2,\delta)$-LRC with parameters $[n,2\rho-2,2\delta]_q$.
\end{theorem}
\begin{IEEEproof}
Let $\xi$ be a primitive $n$-th root of unity. Since $\gcd(\rho,\delta+1)=1$, there is an integers $a$ such that $a\rho\equiv 1(\bmod ~\delta+1)$. Let
 $$g(x)=\prod_{i=1}^{\delta-1}(x^{\rho}-\alpha^{i})(x-\gamma_1)(x-\gamma_2),$$
 where $\alpha=\xi^{\rho}$, $\gamma_1=\alpha^{a\delta}$, $\gamma_2=\alpha^{a(\delta+1)}$.
Since $\alpha^{q-1}=\xi^{\frac{n}{\delta+1}(q-1)}=1$, we have $\alpha^{q}=\alpha$, $\gamma_{1}^{q}=\gamma_1$ and $\gamma_{2}^{q}=\gamma_2$, then $g(x)^q=g(x^q)$, i.e., $g(x)\in \mathbb{F}_{q}[x]$. Furthermore,
$\gamma_1^{\rho}=\alpha^{\delta}$, $\gamma_2^{\rho}=\alpha^{\delta+1}$, so $\gamma_1$, $\gamma_2$ are two extra zeros of $g(x)$ besides $\{\xi^{i+(\delta+1)j}~|~1\leq i \leq\delta-1, 0\leq j\leq \rho-1\}$, i.e., the degree of $g(x)$ is $(\delta-1)\rho+2$. From $\gamma_i^{n}=1$ for $i=1,2$, we have $g(x)$ divides $x^n-1$.

Let $\mathcal{C}$ be a $q$-ary cyclic code of length $n$ with generator polynomial $g(x)$, from Lemma~\ref{lem:locality}, $\mathcal{C}$ has $(2,\delta)$-locality.

Let $i_1=\langle a\rho\delta \rangle_{n}$, $i_2=\langle a\rho(\delta+1)\rangle_{n}$, then $\gamma_1=\xi^{i_1}$, $\gamma_2=\xi^{i_2}$. Since $a\rho\equiv 1(\bmod ~\delta+1)$, $i_1\not\equiv i_2(\bmod ~\delta+1)$, by Corollary~\ref{cor:boubdforlrc} we have $d(\mathcal{C})\geq 2\delta$. On the other hand, according to Corollary~\ref{cor:icc}, $d\leq 2\delta$, so $d=2\delta$, which means $\mathcal{C}$ is distance-optimal.
\end{IEEEproof}

For the case of $(\delta+1)\mid (q+1)$, depending on whether $\delta$ is even or odd, we have the following two constructions of distance-optimal LRCs with unbounded length from cyclic codes and constacyclic codes respectively.

\begin{theorem}\label{thm:olrcq+1deven}
Let $q$ be a prime power and $\delta\geq 2$ be an even integer such that $(\delta+1)\mid (q+1)$. If $\gcd(\rho,\delta+1)= 1$, then there exists a distance-optimal $(2,\delta)$-LRC with parameters $[n,2\rho-2,2\delta]_q$.
\end{theorem}

\begin{IEEEproof}
Let $\xi$ be a primitive $n$-th root of unity. Since $\gcd(\rho,\delta+1)=1$, there is an integers $a$ such that $a\rho\equiv 1(\bmod ~\delta+1)$. Let
 $$g(x)=\prod_{i=-\frac{\delta-2}{2}}^{\frac{\delta-2}{2}}(x^{\rho}-\alpha^{i})(x-\gamma_1)(x-\gamma_2),$$
 where $\alpha=\xi^{\rho}$, $\gamma_1=\alpha^{\frac{a\delta}{2}}$, $\gamma_2=\alpha^{-\frac{a\delta}{2}}$.
Since $\alpha^{q+1}=\xi^{\frac{n}{\delta+1}(q+1)}=1$, we have $\alpha^{q}=\alpha^{-1}$, $\gamma_{1}^{q}=\gamma_{2}$ and $\gamma_{2}^{q}=\gamma_{1}$, then $g(x)^q=g(x^q)$, i.e., $g(x)\in \mathbb{F}_{q}[x]$. Furthermore,
$\gamma_1^{\rho}=\alpha^{\frac{\delta}{2}}$, $\gamma_2^{\rho}=\alpha^{-\frac{\delta}{2}}$, so $\gamma_1$, $\gamma_2$ are two extra zeros of $g(x)$ besides $\{\xi^{i+(\delta+1)j}~|~-\frac{\delta-2}{2}\leq i \leq\frac{\delta-2}{2}, 0\leq j\leq \rho-1\}$, i.e., the degree of $g(x)$ is $(\delta-1)\rho+2$.
 From $\gamma_i^{n}=1$ for $i=1,2$, we have $g(x)$ divides $x^n-1$.

Let $\mathcal{C}$ be a $q$-ary cyclic code of length $n$ with generator polynomial $g(x)$, from Lemma~\ref{lem:locality}, $\mathcal{C}$ has $(2,\delta)$-locality.

Let $i_1=\langle \frac{a\rho\delta}{2}\rangle_{n}$, $i_2=\langle -\frac{a\rho\delta}{2}\rangle_{n}$, then $\gamma_1=\xi^{i_1}$, $\gamma_2=\xi^{i_2}$. Since $a\rho\equiv 1(\bmod ~\delta+1)$, $i_1\not\equiv i_2(\bmod ~\delta+1)$, by Corollary~\ref{cor:boubdforlrc} we have $d(\mathcal{C})\geq 2\delta$. On the other hand, according to Corollary~\ref{cor:icc}, $d\leq 2\delta$, so $d=2\delta$, which means $\mathcal{C}$ is distance-optimal.
\end{IEEEproof}

\begin{theorem}\label{thm:olrcq+1dodd}
Let $q$ be a prime power and $\delta\geq 2$ be an odd integer such that $(\delta+1)\mid (q+1)$. If $\gcd(\rho,\delta+1)= 1$, then there exists a distance-optimal $(2,\delta)$-LRC with parameters $[n,2\rho-2,2\delta]_q$.
\end{theorem}

\begin{IEEEproof}
Since $\delta+1$ is even and $\gcd(\rho, \delta+1) = 1$, $\rho$ is odd. Note that $q$ is odd, then there is a unique positive
integer $\mu$ such that $2^{\mu}\mid(q -1)$ and $2^{\mu+1}\nmid(q-1)$. Clearly,
$\gcd(\rho, 2^{\mu}(\delta+1)) = 1$, so there is an integer $a$ such that
$a\rho\equiv 1 (\bmod ~2^{\mu}(\delta+1))$.

Let
 $$g(x)=\prod_{i=\frac{\frac{q-1}{\kappa}-(\delta-2)}{2}}^{\frac{\frac{q-1}{\kappa}+(\delta-2)}{2}}(x^{\rho}-\eta^{\rho} \alpha^{i})(x-\gamma_1)(x-\gamma_2),$$
where $\kappa = 2^{\mu}$, $\eta$ is a primitive $\kappa n$-th root of unity, $\xi=\eta^{\kappa}$, $\alpha=\xi^{\rho}$, $\gamma_1=\eta^{a\rho}\alpha^{a\frac{\frac{q-1}{\kappa}+\delta}{2}}$,
$\gamma_2=\eta^{a\rho}\alpha^{a\frac{\frac{q-1}{\kappa}-\delta}{2}}$.

We can check that $(\eta^{\rho} \alpha^{i})^{q}=\eta^{\rho}\eta^{\rho(q-1)} \alpha^{-i}=\eta^{\rho} \alpha^{(\frac{q-1}{\kappa}-i)}$ for all $i \in \mathbb{Z}$, from which we have $$\gamma_1^{q}=((\eta^{\rho}\alpha^{\frac{\frac{q-1}{\kappa}+\delta}{2}})^{q})^{a}=(\eta^{\rho}\alpha^{\frac{\frac{q-1}{\kappa}-\delta}{2}})^{a}=\gamma_2,$$
and similarly $\gamma_2^{q}=\gamma_1$.
So $g(x)^q = g(x^q)$, i.e., $g(x) \in \mathbb{F}_q[x]$.
Furthermore, $\gamma_1^{\rho}=\eta^{\rho}\alpha^{\frac{\frac{q-1}{\kappa}+\delta}{2}}$,
$\gamma_2^{\rho}=\eta^{\rho}\alpha^{\frac{\frac{q-1}{\kappa}-\delta}{2}}$, which means that $\gamma_1$ and $\gamma_2$ are indeed two extra zeros of $g(x)$ besides $\{\eta\xi^{i+(\delta+1)j}~|~\frac{\frac{q-1}{\kappa}-(\delta-2)}{2}\leq i \leq \frac{\frac{q-1}{\kappa}+(\delta-2)}{2}, 0\leq j\leq \rho-1\}$, i.e., the degree of $g(x)$ is $(\delta-1)\rho+2$.
 From $\gamma_i^{n}=\eta^{a\rho n}=\eta^{n}\eta^{(a\rho-1)n}=\eta^{n}$ for $i=1,2$, we have $g(x)$ divides $x^n-\eta^{n}$.

Let $\mathcal{C}$ be a $q$-ary $\eta^{n}$-constacyclic code of length $n$ with generator polynomial $g(x)$, from Lemma~\ref{lem:locality}, $\mathcal{C}$ has $(2,\delta)$-locality.

As $a\rho\equiv 1(\bmod ~2^{\mu}(\delta+1))$, let $a\rho-1=m\kappa(\delta+1)$, then
$\gamma_1=\eta^{a\rho}\alpha^{a\frac{\frac{q-1}{\kappa}+\delta}{2}}=\eta\xi^{m(\delta+1)+a\rho\frac{\frac{q-1}{\kappa}+\delta}{2}}$,
$\gamma_2=\eta^{a\rho}\alpha^{a\frac{\frac{q-1}{\kappa}-\delta}{2}}=\eta\xi^{m(\delta+1)+a\rho\frac{\frac{q-1}{\kappa}-\delta}{2}}$.
Let $i_1=\langle m(\delta+1)+a\rho\frac{\frac{q-1}{\kappa}+\delta}{2}\rangle_n$, $i_2=\langle m(\delta+1)+a\rho\frac{\frac{q-1}{\kappa}-\delta}{2}\rangle_n$, then $\gamma_1=\eta\xi^{i_1}$, $\gamma_2=\eta\xi^{i_2}$. Since $a\rho\equiv 1(\bmod ~\kappa(\delta+1))$, $i_1\not\equiv i_2(\bmod ~\delta+1)$, by Proposition~\ref{prop:boundforconsta} and Corollary~\ref{cor:boubdforlrc} we have $d(\mathcal{C})\geq 2\delta$. On the other hand, according to Corollary~\ref{cor:icc}, $d\leq 2\delta$, so $d=2\delta$, which means $\mathcal{C}$ is distance-optimal.
\end{IEEEproof}

\begin{remark}
The lengths of distance-optimal LRCs constructed in Theorems~\ref{thm:olrcq-1}, \ref{thm:olrcq+1deven}, and \ref{thm:olrcq+1dodd} have the form $n=\rho(\delta+1)$, where $\rho$ is independent of $q$. Therefore, for a fixed field size $q$, the code length $n$ can be arbitrarily large.
It is worth noting that in the works of \cite{Cai20}, \cite{Bocong19}, \cite{Luo19}, \cite{Luo23}, \cite{Fang20}, \cite{Sun19}, the constructions of Singleton-optimal (distance-optimal) $(r, \delta)$-LRCs with unbounded length consistently yield the minimum distance not exceeding $r+\delta-1$ (the reader may refer to TABLE III in \cite{Luo23} for the detailed comparison of parameters). In Theorems~\ref{thm:olrcq-1}, \ref{thm:olrcq+1deven}, and \ref{thm:olrcq+1dodd} of this paper, for cases where $\delta\geq 2$, we observe that the minimum distance $d=2\delta> \delta+1=r+\delta-1$.
\end{remark}

\section{Conclusion}
In this paper, we investigated the lower bounds for the minimum distance of cyclic codes using the singleton procedure technique. We have provided a bound for the minimum distance of cyclic codes with a special type of defining set. This bound, in turn, serves as a foundation for deriving numerous new bounds in explicit forms. Specially, three new bounds with general patterns are provided, one of which can serve as a generalization of the Betti-Sala bound. Furthermore, for the case of $s=1$, we extend these new bounds, which enable the new constructions of cyclic and constacyclic $(2,\delta)$-LRCs with unbounded length. It turns out that the new LRCs are distance-optimal, and to the best of our knowledge, are the first class of distance-optimal $(r,\delta)$-LRCs with unbounded length and minimum distance exceeding $r+\delta-1$.

As for future work, it is intriguing to continue constructing additional optimal cyclic and constacyclic $(r,\delta)$-LRCs by utilizing the new bounds.

%

\end{document}